\documentclass[aps,prx,twocolumn,superscriptaddress,preprintnumbers]{revtex4-2}

\usepackage[T1]{fontenc}
\usepackage{amssymb}
\usepackage{graphicx}
\usepackage{color}
\usepackage{caption}
\usepackage{hyperref}
\usepackage{amsthm}
\usepackage{amsmath}
\usepackage{braket}
\usepackage{appendix}
\usepackage{subcaption}
\usepackage{comment}

\usepackage{tikz}
\usetikzlibrary{patterns}
\usepackage{physics}

\usepackage{amsthm}
\usepackage{enumitem}
\usepackage{comment}

\begin{document}


\title{
Frustrating quantum batteries 
}

\author{A. G. Catalano}
\affiliation{Institut Ru\dj er Bo\v{s}kovi\'c, Bijeni\v{c}ka cesta 54, 10000 Zagreb, Croatia}
\affiliation{Universit\'e de Strasbourg, 4 Rue Blaise Pascal, 67081 Strasbourg, France}

\author{S. M. Giampaolo}
\affiliation{Institut Ru\dj er Bo\v{s}kovi\'c, Bijeni\v{c}ka cesta 54, 10000 Zagreb, Croatia}

\author{O. Morsch}
\affiliation{CNR-INO and Dipartimento di Fisica dell’Universit\`a di Pisa, Largo Pontecorvo 3, 56127 Pisa, Italy}

\author{V. Giovannetti}
\affiliation{NEST, Scuola Normale Superiore and Istituto Nanoscienze-CNR, I-56126 Pisa, Italy}

\author{F. Franchini}
\email{fabio@irb.hr}
\affiliation{Institut Ru\dj er Bo\v{s}kovi\'c, Bijeni\v{c}ka cesta 54, 10000 Zagreb, Croatia}
\preprint{RBI-ThPhys-2023-02}


\begin{abstract}
{We propose to use a quantum spin chain as a device to store and release energy coherently (namely, a quantum battery) and we investigate the interplay between its internal correlations and outside decoherence. 
We employ the quantum Ising chain in a transverse field, and our charging protocol consists of a sudden global quantum quench in the external field to take the system out of equilibrium. 
Interactions with the environment and decoherence phenomena can dissipate part of the work that the chain can supply after being charged, measured by the ergotropy. 
We find that the system shows overall remarkably better performances, in terms of resilience, charging time, and energy storage, when topological frustration is introduced by setting AFM interactions with an odd number of sites and periodic boundary conditions.
Moreover, we show that in a simple discharging protocol to an external spin, only the frustrated chain can transfer work and not just heat.}
\end{abstract}

\maketitle

\section{Introduction}

In recent years, there has been a global surge of interest in harnessing quantum phenomena at the microscopic level, driven by the rapid advancement of new quantum technologies~\cite{Acin2018, Riedel2017}. 
Within this context, an intriguing area of exploration is the study of "quantum batteries"~\cite{Alicki2013, Binder2015, Campaioli2017, Ferraro2018, Andolina2018, Andolina2019b, Farina2019, Hovh2013}, which are quantum mechanical systems designed for energy storage. 
Quantum batteries (QBs) utilize quantum effects to achieve more efficient and rapid charging processes compared to classical systems. 
This burgeoning field of research encompasses numerous intriguing questions, ranging from the stabilization of stored energy~\cite{Ghera2020, Rosa2020} to the investigation of optimal charging protocols~\cite{Tirone2021, Tirone2022, Tirone2023a, Tirone2023b, Rodri2022, Pirmo2019, Mazzoncini2022, Erdman2022}.
One of the first practical implementations of this type of device is the quantum Dicke battery in  
Ref.~\cite{Ferraro2018} where the energy from a photonic cavity mode (acting as a charger) is transferred to a battery comprising $N$ quantum units, each described by a two-level system. 
Such a model exhibits a collective speed-up~\cite{Andolina2019a} in the charging process.
The Dicke battery has garnered significant interest due to its versatility in various implementation platforms (e.g., superconducting qubits~\cite{Wang2020}, quantum dots~\cite{Stock2017, Samk2018}, coupled with a microwave resonator, Rydberg atoms in a cavity~\cite{Haroche2012}, etc.), leading to the exploration of numerous variations of this model~\cite{Zhang2019, Crescente2020a, Crescente2020b, Crescente2020c, Dou2022a, Dou2022b, Zhao2022, Rossini2019, Rossini2020, Quach2022}.   

To have a practical application, QBs must however not only rapidly store energy but also be able to provide useful energy (i.e. work) once charged~\cite{Andolina2019b, Monsel2020, Maffei2021, Oppe2002}. 
A crucial aspect of the problem is to assess the stability of these models in realistic scenarios where they are {subject to} environmental noise.
Preliminary studies in this direction have been
obtained in Refs.~\cite{Carrega2020, Bai2020, Tabesh2020, Ghosh2021, Santos2021, Zaka2021, Landi2021, Morrone2023, Sen2023, Liu2019, Santos2019, Santos2020, Quach2020, Santos2020, Ghera2020, Liu2021, Arjmandi2022} where various schemes have been proposed to stabilize QBs in the presence of specific types of perturbations.
In Refs.~\cite{Tirone2021, Tirone2023a, Tirone2023b, Tirone2022} a general theory of work extraction for noisy QB models composed by large collections of non-interacting subsystems (quantum cells) have been presented. 
The fundamental theoretical tool for this type of study is provided by the {\it ergotropy}~\cite{Alla2004, Niedenzu2019}, a non-linear functional that gauges the maximum amount of energy that can be extracted from an assigned input state of a quantum system under reversible, i.e. unitary operations that do not alter the system entropy. 

{ The present paper contributes to the development of the theory of QBs.
Here, we analyze the work extraction from QBs made of $N$ interacting spins which, once charged, undergo complete dephasing in the energy eigenbasis of the associated Hamiltonian. 
More precisely, our analysis is focused on many-body models which exhibit exotic behaviors when proper Frustrated Boundary Conditions (FBCs) are imposed.}
A typical example is represented by a linear chain of an odd number of spins arranged in a ring geometry (i.e. with periodic boundary conditions): when {classically} paired with antiferromagnetic (AFM) interactions, such a system cannot realize the perfect Neel ordering, hence exhibiting topological frustration due to the presence of a ferromagnetic (FM) kink along the chain. 
At the quantum level, the introduction of such frustration radically modifies the structure of both the ground-state manifold~\cite{Maric2020_induced, Maric2020_destroy, Catalano2022} and of the low-energy spectrum~\cite{Maric2022_fate}, leading to a whole set of novel behaviors~\cite{Torre2022, Maric2022_nature, Giampaolo2019, Odavic2022} which are potentially interesting for technological applications. 
One important example is that while in non-frustrated models (at least far from critical points) the ground state manifold is separated by a finite energy gap from the rest of the spectrum, for the topologically frustrated systems it belongs to a band (for the ring geometry discussed above the gap closes as $N^{-2}$). 

{{As a charging mechanism we consider a simple (relatively easy to implement)  global quantum quench. Moreover, we show how topological frustration {enhances} the robustness to decoherence of a quantum battery: while in the non-frustrated case the ergotropy of the battery can be reduced to less than $30\%$ of its initial value by decoherence phenomena, we observe that a frustrated battery manages to retain more than $90\%$ of the original ergotropy in all the parameter ranges analyzed. Finally, we propose a simple discharging protocol that shows how it is possible to transfer energy from a many-body quantum battery charged with our protocol to an ancillary spin. Surprisingly, we observe that only frustrated batteries can transfer work efficiently, in the form of ergotropy, while the non-frustrated battery only manages to heat up the ancillary system.}

{The manuscript is organized as follows:
in Sec.~\ref{sec:model} we introduce the quantum spin models and the charging protocol that we consider to realize a quantum battery, as well as introducing the role of decoherence in these systems. In Sec.~\ref{sec:FUQB} we compare the performances of frustrated and non-frustrated batteries under the assumption of \textit{fast charging}, i.e. considering a purely coherent charging protocol. In Sec.~\ref{sec:deco-charg} we drop this assumption and analyze the effect of decoherence during the charging protocol introducing a non-unitary dynamics during the quantum quench. 
In Sec.~\ref{sec:discharge} we present a protocol for energy transfer from a many-body quantum battery to a single ancillary spin. Finally, we discuss our results and possible developments in Sec.~\ref{sec:conclusion}.  
}}
 
\begin{figure}[t]
\includegraphics[width=0.48\textwidth]{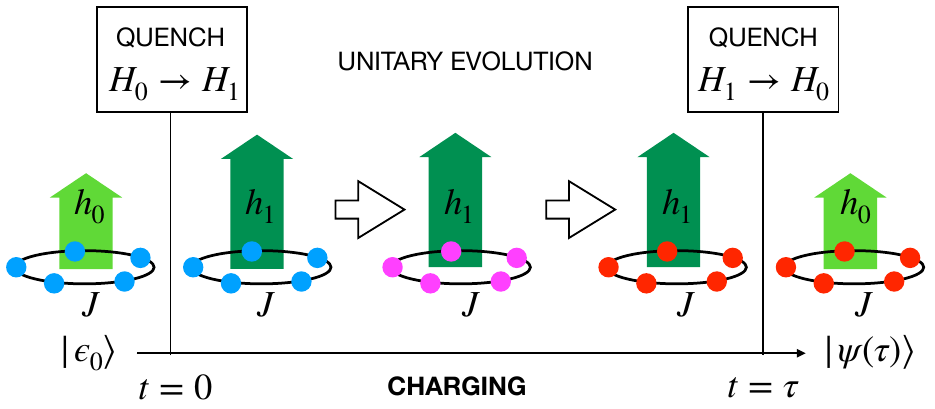}
\caption{Schematic overview of the charging process: the QB is represented by a collection of an odd number $N$ of spin-1/2 particles, initialized in the ground state $|\epsilon_0\rangle$ of the Ising Hamiltonian $H_0=H(J,h_0)$ with local field $h=h_0$ and {coupling $J$} whose modulus is equal to $1$. 
Setting $J=-1$ corresponds to considering a non-frustrated, FM QB model while setting $J=1$  to a frustrated, AFM QB.
Energy is pumped into the system in the time interval $[0,\tau]$ via the unitary evolution associated with the Hamiltonian $H_1=H(J,h_1)$ realized by quenching the local field from $h_0$ to $h_1$ at time $t=0$ and
by restoring it to the value $h_1$ at $t=\tau$.}
\label{fig:scheme}
\end{figure}

\section{The theoretical framework}\label{sec:model} 
 
\subsection{The model}

While the phenomenology of topological frustration has already been described in detail for more general models like the $XYZ$ chain~\cite{Catalano2022}, without losing generality, {here} we will focus on the simplest case, i.e. a ring of an odd number $N$ of spin-1/2 particles coupled via the quantum Ising chain in a transverse magnetic field. 
The Hamiltonian of such a model is 
\begin{equation}
\label{eq:Ising}
H(J,h)= J\sum_{j=1}^N \sigma^x_{l}\sigma^x_{l+1}-h\sum_{l=1}^N\sigma^z_{l} \;,
\end{equation}
where $\sigma_l^{\alpha}$ ($\alpha=x,\,y,\,z$) are  Pauli operators acting on the $l$-th spin,   $\sigma_{N+1}^{\alpha}= \sigma_1^{\alpha}$ enforces the period boundary conditions, and $h$ is the strength of the external field.
The constant $J$ governs the nature of the couplings among the spins: its modulus $|J|$  determines the strength of the Ising interactions and in our analysis will be fixed equal to $1$, while its sign allows us to tune from an AFM frustrated system, hence frustrated for $J>0$, to an non-frustrated FM one for $J<0$.  

Regardless of the sign of $J$, the model is analytically integrable, and a detailed solution can be found in Appendix~\ref{ap:Ising}. 
Thanks to this, it is possible to observe how while some properties of the system are not affected by the presence or absence of topological frustration, others assume very different behaviors.
An example of the latter is the existence of an energy gap between the ground state manifold and the closest set of excited states in the ordered phase $|h|<1$. 
If, in the non-frustrated case, in the thermodynamic limit the two-fold ground state manifold is separated from the band of excited states by a finite energy gap equal to $\Delta_{FM}=1-|h|$, {this} disappears completely in presence of frustration. 
In fact, for $J=1$, at a finite size, the ground state is part of a band made of $2N$ states in which the gap between the elements with lower energy closes according to the law
\begin{equation}
  \label{eq:gap_afm}
\Delta_{AFM}=\frac{2|h|}{1-|h|}\frac{\pi^2}{N^2}+O(N^{-4}).  
\end{equation}
Hence, the frustrated AFM model presents a gap that vanishes quadratically with the system's size. 
It is worth noting that, fixing $N$ to an odd integer number and moving towards the critical points $h_c=\pm 1$, the gap of the non-frustrated model decreases while that of the frustrated one increases. 


\subsection{The charging protocol}

To store energy in a spin system as the one described in \eqref{eq:Ising}, i.e. to use such a system as a QB, we propose a simple protocol based on quenches of the external magnetic field sketched in Fig.~\ref{fig:scheme}.  
The starting point is the ground state $\ket{\epsilon_0}$ with associated energy $\epsilon_0$ of the Hamiltonian $H_0=H(J,h_0)$.
Such Hamiltonian admits a set of eigenstates that we denoted as $\{ \ket{\epsilon_\ell}\}$, ordered in such a way that the associated eigenvalues $\epsilon_l$ satisfy the following condition $\epsilon_\ell\le\epsilon_{\ell+1}$. 
At time $t=0$, we perform a sudden global quench to the Hamiltonian $H_1=H(J,h_1)$, whose eigenstates we denote by $\{ \ket{\mu_k}\}$,
ordered in such a way that $\mu_k\le \mu_{k+1}$. 
The system then evolves unitarily under the action of $H_1$ for a certain time interval $\tau$ at which the system Hamiltonian is quenched back to $H_0$ to close the charging cycle. {Note that $h_1$ can also be greater than $|J|=1$ crossing the Ising quantum critical point and thus the charging process can also happen in a different phase, before we return to $h_0$.}
In the absence of external interferences, the QB at the end of the charging process is described by the vector $|\psi(\tau)\rangle= e^{-\imath H_1\tau} |\epsilon_0\rangle$, and the energy stored is given by
\begin{equation}\label{Ein} 
 E_{in}=\langle \psi(\tau)| H_0 |\psi(\tau)\rangle-\epsilon_0  = 
 \sum_{\ell} P_\ell(\tau)  (\epsilon_\ell - \epsilon_0)\;,
\end{equation}
where the populations $P_\ell(\tau)$ are
\begin{eqnarray} P_\ell(\tau) = \big|\sum_{k} e^{-\imath  \mu_k \tau}  \bra{\mu_k}\ket{\epsilon_0} \bra{\epsilon_\ell}\ket{\mu_k}\label{popu0} 
\big|^2\;. 
\end{eqnarray} 
Thanks to the integrability of the Hamiltonian in \eqref{eq:Ising}, it is possible to derive analytically the populations $P^{(0)}_\ell(\tau)$, see Appendix~\ref{ap:population} for details.
In Fig.~\ref{fig:inversion2} we have plotted the $P^{(0)}_\ell(\tau)$ for both the frustrated and non-frustrated case for a specific choice of the system parameters. 
From the figure, it is possible to observe that, at the level of the populations of the eigenstates of $H_0$, there is no clear difference between the two cases. 
As a consequence, also the amount of energy stored in the system is almost the same with small differences that vanish {by} increasing the system size.

\begin{figure}[t!]
\includegraphics[width=.99\columnwidth]{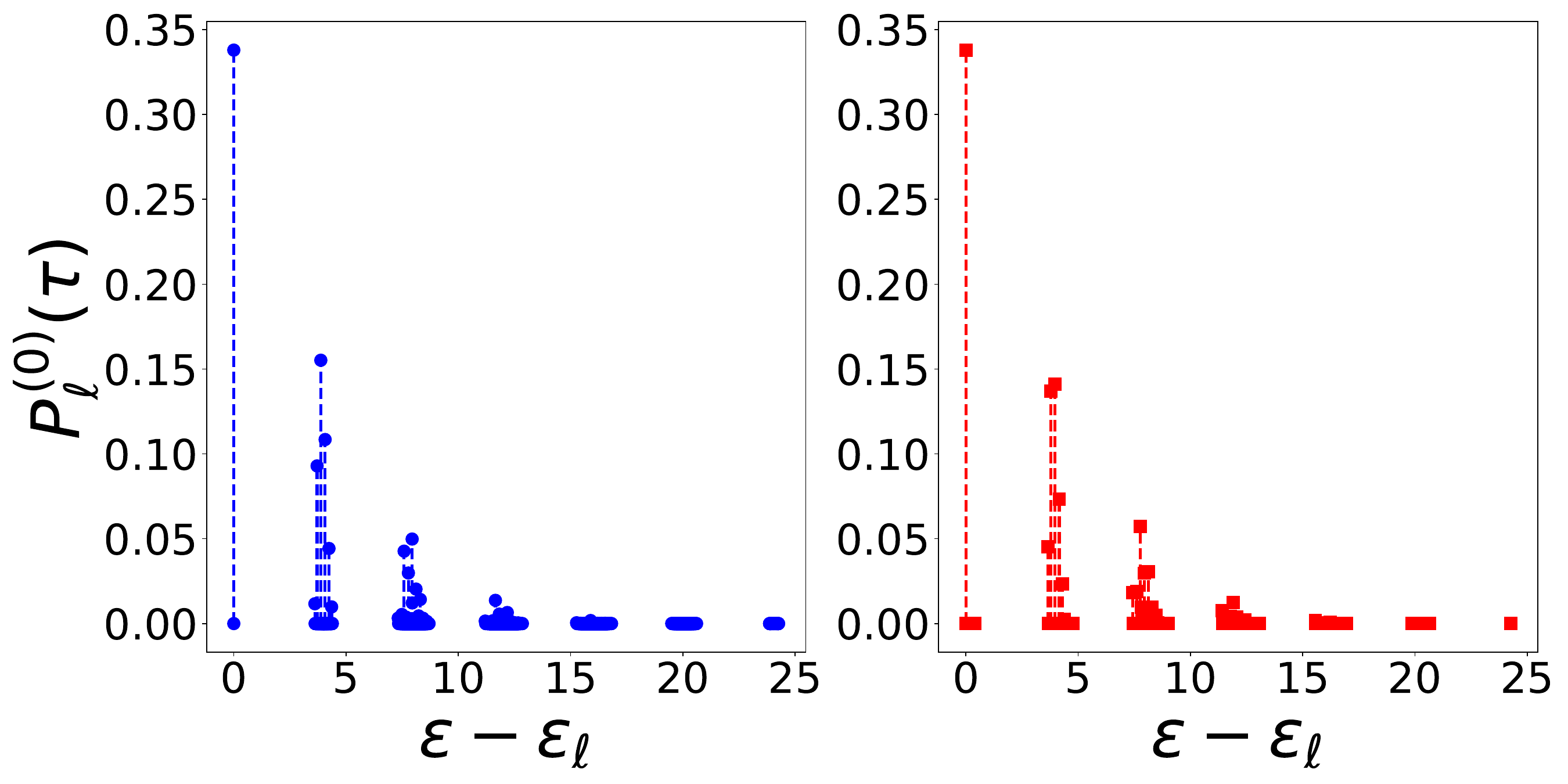}
\caption{\small Plot of the distribution $ P_\ell(\tau)$ of Eq.~(\ref{popu0})   which define the populations of the energy eigenspaces of the QB Hamiltonian $H_0=H(J,h_0)$ after a cyclic quench from $h_0=0.1$ to $h_1=0.8$ and back to $h_0$, for the frustrated $J=1$ (right panel) and non-frustrated $J=-1$ (left panel) systems. This is obtained for a chain of $N=13$ spins setting $\tau=1.0$.}
\label{fig:inversion2}
\end{figure}

\subsection{The role of decoherence} 

{As long as the evolution of the system remains unitary, it is always possible to reverse it, hence completely recovering the stored energy $E_{in}$.
But, in presence of decoherence, the dynamics of a quantum system becomes non-unitary, and hence there is no unitary transformation that can bring {the system back} to its initial state, thereby reducing the amount of energy that can be extracted from it~\cite{Alla2004, Niedenzu2019}. 
However, one of the main problems in the study of decoherence is that the results obtained are, in general, strongly dependent on the non-unitary dynamics model taken into account which, in turn, depends on the specifics of the experimental apparatus in which the model is tested. 
Since we aim to carry out a theoretical analysis as independent as possible from the possible experimental realization, we have decided to consider, as a source of decoherence, a purely dephasing dynamics such as the one induced by the master equation proposed by Milburn~\cite{Milburn1991}}
\begin{equation}
\label{eq:milburn}
\dot{\rho}(t)=-\imath [H(t),\rho(t)]-\frac{1}{2\nu}[H(t),[H(t),\rho(t)]].
\end{equation}
Here $\rho(t)$ and $H(t)$ are the instantaneous system density matrix and Hamiltonian, $\dot{\rho}(t)$ is the derivative of the density matrix and $\nu$ parametrizes the characteristic decoherence rate of the model.

\begin{figure}[t!]
\includegraphics[width=.99\columnwidth]{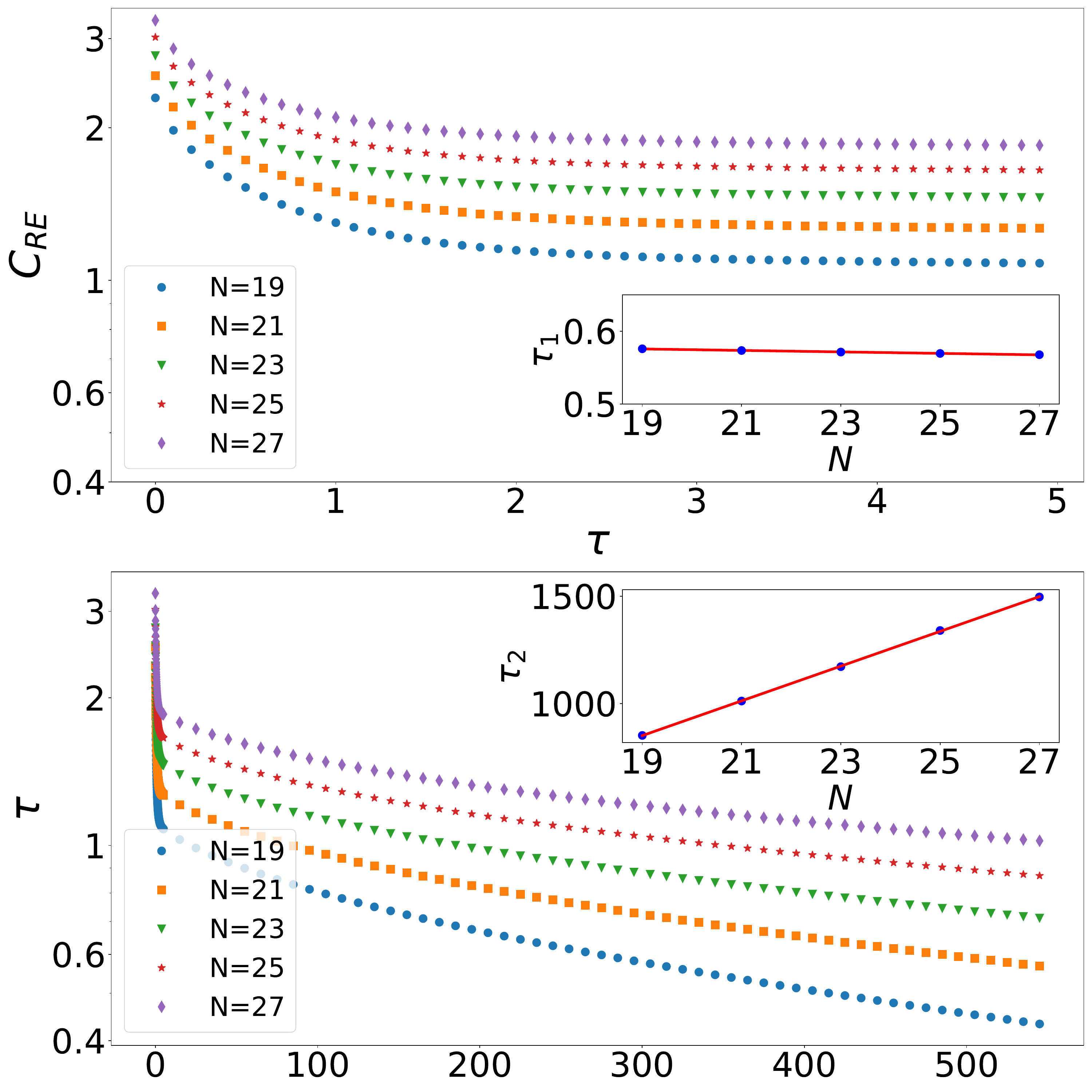}
\caption{\small {Plotting the entropy of coherence in Eq. \eqref{eq:quantum_coherence} as a function of time we observe the emergence of two distinct decoherence time scales. A fast one, characterized by a short time $\tau_1$ (top panel and inset), which is approximately independent of $N$, and a slower one with characteristic time $\tau_2\gg \tau_1$ (bottom panel and inset), which instead is proprotional to $N$. Data have been obtained for $J=1$, $h_0=0.1$, $\Delta h=0.5$. The results are the same for the frustrated and non-frustrated systems. Quantitatively, from a best fit analysis we find that $\tau_1\approx 0.05 \frac{\nu}{(J-h_0)^2}+10^{-3} N$, while $\tau_2\approx81 N -688$.} }
\label{fig:time_scales}
\end{figure}

{ Taking into account the charging process that we have introduced, and hence the dependence on time of the Hamiltonian, the second term of the r.h.s. of \eqref{eq:milburn} implies that all the off-diagonal terms of the matrix are exponentially suppressed {in the energy eigenbasis} with a characteristic decoherence time equal to $\tau_{k,l}\approx\frac{2\nu}{\Delta E_{k,l}^2}$, where $\Delta E_{k,l}$ is the energy difference between the states $\ket{\epsilon_k}$ $\ket{\epsilon_l}$ (or $\ket{\mu_k}$ $\ket{\mu_l}$ {when we will consider a slow charging process.}).
At this point, it is important to note that the global quench $H_0\leftrightarrow H_1$ preserves all the symmetries of the Hamiltonian, most importantly the translational and the parity symmetry.  
Therefore, since the initial state $|{\epsilon_0}\rangle$ {is an eigenstate of the momentum operator with zero momentum and fixed parity}, then the states $\epsilon_{\ell}$ with $P_{\ell}(\tau)\neq 0$ are also eigenstates with the same parity as the ground state and vanishing momentum~\cite{Torre2022}.
{Each of these states is associated with a different energy and hence, for such states, $\Delta E_{k,l}$ always differs from zero.}
As a consequence, after a sufficiently long time, due to the effect of the non-unitary dynamics, the state of the QB will be well approximated by a diagonal density matrix with zero coherence in the eigenbasis of the Hamiltonian.}
To understand how large the characteristic times of the {system's dephasing} are, from Fig.~\ref{fig:inversion2} we can see that states with a non-vanishing population belong to different energy bands. 
Hence, if the off-diagonal term $\rho_{k,l}$ is generated from two states coming from two different bands, the timescale of its exponential suppression will be proportional to the {\em fast decoherence time} {$\tau_1$, which, by dimensional analysis, we expect of the order of $\nu/(J-h)^2$ and independent of $N$.}
On the contrary, if the two states $\ket{\epsilon_k}$ and $\ket{\epsilon_l}$ belong to the same energy band, such timescale will be related to the {\em slow decoherence time} to {$\tau_2\gg\tau_1$, which depends on the average intraband level energy spacing between occupied levels and turns out to be proportional to the system's size $N$.} 
The existence of two different timescales in the non-unitary dynamics induced by eq.~\eqref{eq:milburn} can be appreciated by looking at Fig. \ref{fig:time_scales} in which, we have depicted the behavior of the relative entropy of coherence for the state $\rho(t)$, i.e. the $C_{RE}(\rho(t))$~\cite{Baumgratz2014}, {and provided an estimate of $\tau_{1,2}$ for some parameter choice.}
The relative entropy of coherence is defined as 
\begin{equation}
\label{eq:quantum_coherence} 
C_{RE}(\rho(t))=S(\rho_D(t))-S(\rho(t))
\end{equation}
{where $S(x)=-\sum_i \lambda_i \log \lambda_i$ is the von Neumann entropy of the density matrix $x$ with eigenvalues $\{\lambda_i\}$ and $\rho_D(t)$ is the diagonal matrix obtained by $\rho(t)$ artificially suppressing all the off-diagonal element in a given basis (the Hamiltonian eigenbasis in our case).} 
From the plot, it is easy to see the existence of two very different time scales. 

The estimation of these times allows for identifying different operating regimes for the QB.
{Since, ideally, a QB should be able to store energy for a long time}, we have decided that, in this article, we will focus on the worst-case scenario, i.e. one where you try to extract work after a time $T \gg \tau_2$ has passed since the end of the charging process {and we leave a detailed analysis of the time-scales $\tau_{1,2}$ and of the behaviors for intermediate times for further works.}
In this {long-time} scenario, the decoherence leads to the complete collapse of the QB density matrix into a diagonal ensemble in the system's energy eigenbasis. 
On the other hand, regarding the charging process, we will specifically examine two distinct charging regimes. 
The first of these is the so-called {\it fast-charging regime}, in which the charging time $\tau$ is considered to be much faster than that of any decoherence time  $\tau\ll\tau_1$. 
As a consequence, the charging process can be considered a unitary process.
On the other hand, the {\it slow-charging regime}, in which $\tau_1$ and $\tau$  are comparable, a partial dephasing occurs also during the charging process.

\section{Batteries in the fast-charging regime}\label{sec:FUQB} 

In the fast-charging regime (i.e. for $\tau_1 \gg \tau$) we can neglect the effect of the dephasing during the charging process. 
Under this hypothesis, the asymptotic state of the QB which emerges from Eq.~(\ref{eq:milburn}) at time $T\gg \tau_2$ corresponds to the completely incoherent (in the basis of the eigenstates of $H_0$) diagonal density matrix state 
  \begin{eqnarray}\label{diagorho}
 \rho(T) =  \sum_{\ell} P_{\ell}(\tau) |\epsilon_\ell\rangle \langle \epsilon_{\ell}|\;,
 \end{eqnarray} 
where $P_{\ell}(\tau)$ are the populations defined in Eq.~(\ref{popu0}).
\begin{figure}[t!]
\centering
\includegraphics[width=.9\columnwidth]{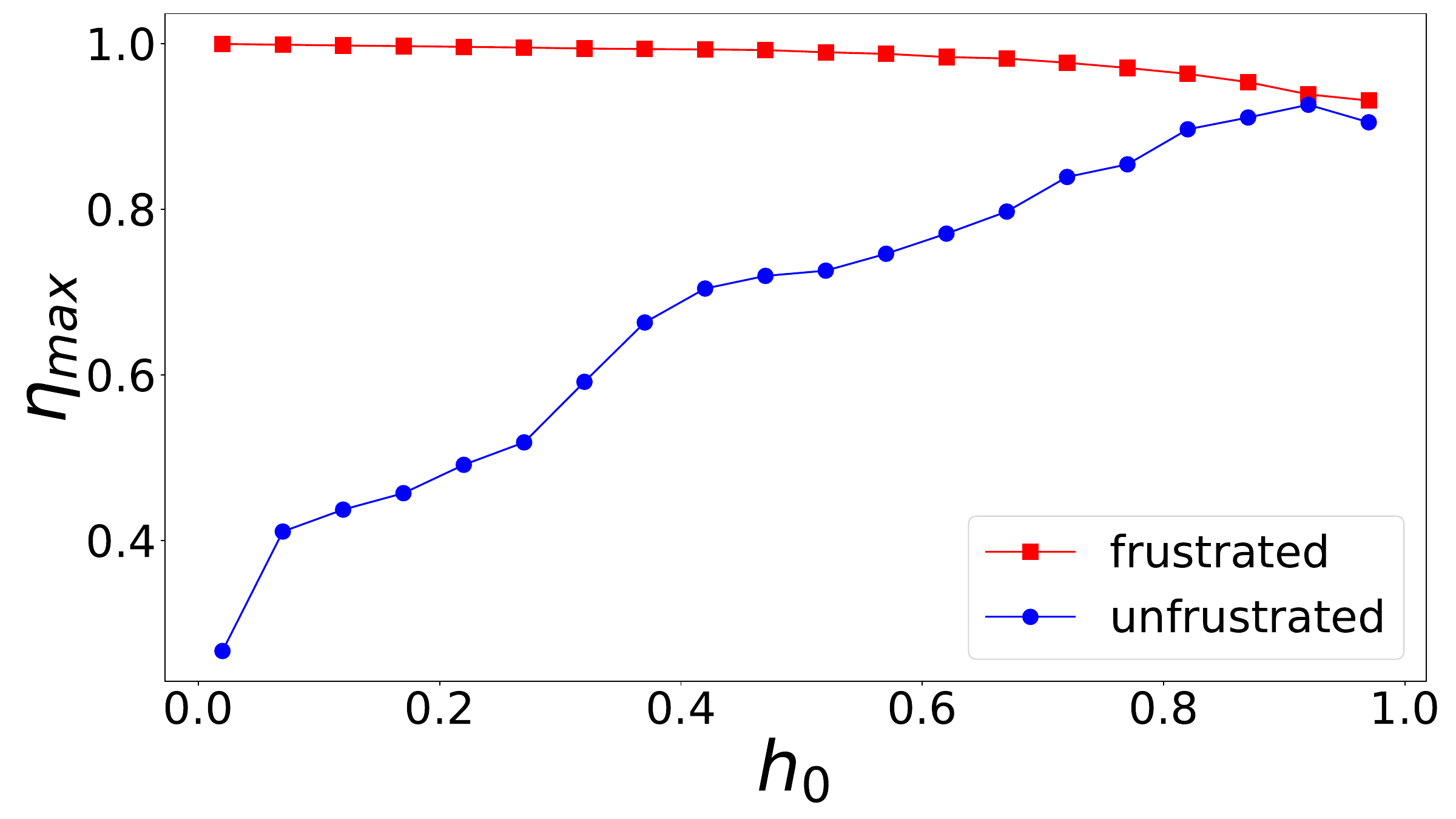}
\includegraphics[width=.9\columnwidth]{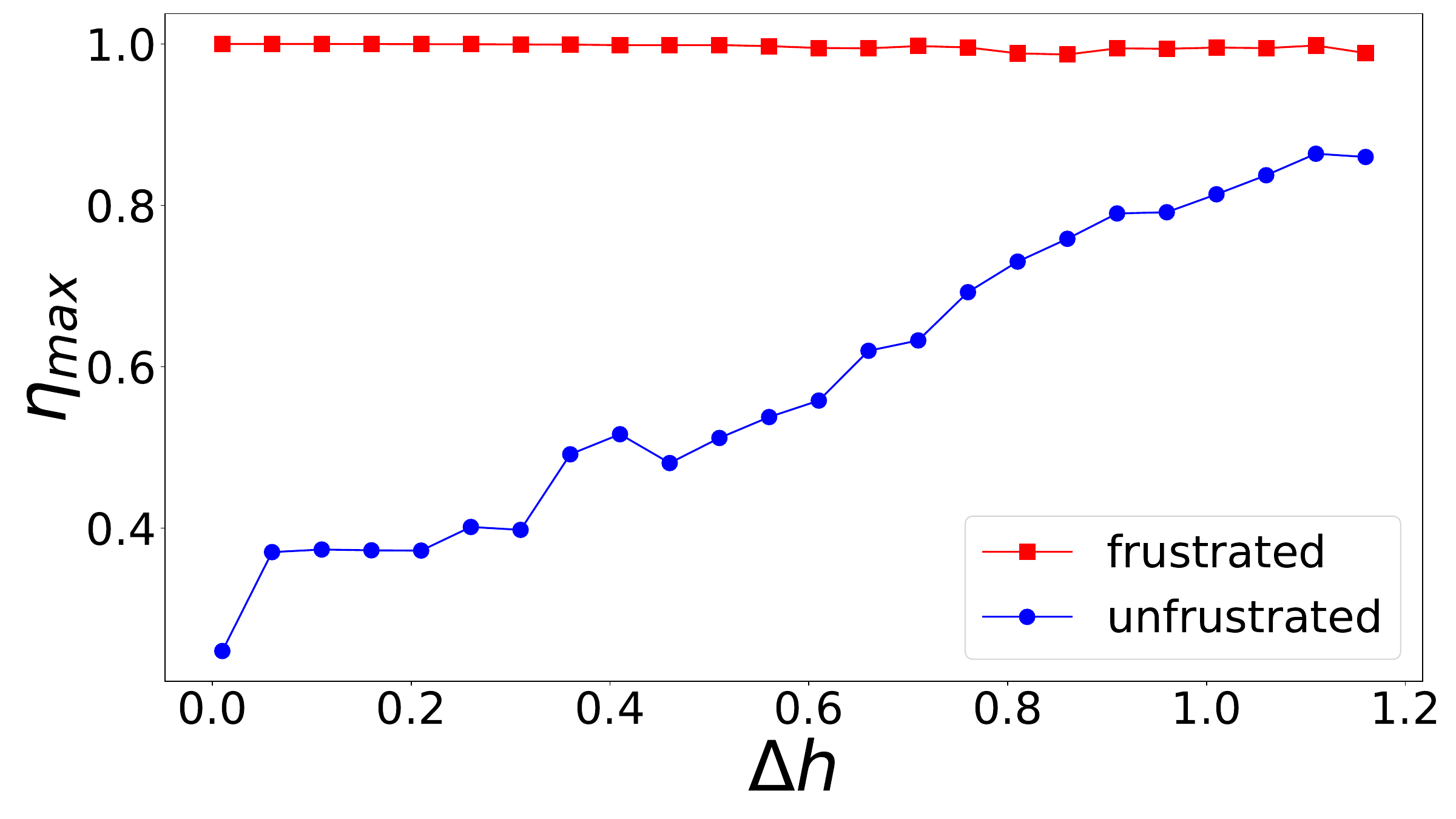}
\caption{\small Maximum value of $\eta$ {defined in Eq. \eqref{eq:eta}}, computed as a function of $h_0$ (upper panel) for $\Delta h=0.01$, and as a function of $\Delta h$ (lower panel) for $h_0=0.001$. Data are obtained for charging times $\tau \in (0,50)$, for a chain of $N=25$ spins. {These plots show how, after decoherence, the frustrated chain has retained most of its charge, while the non-frustrated one typically loses the majority of the initial charge.}}
\label{fig:eta_h}
\end{figure}

\begin{figure}[t!]
\centering
\includegraphics[width=.9\columnwidth]{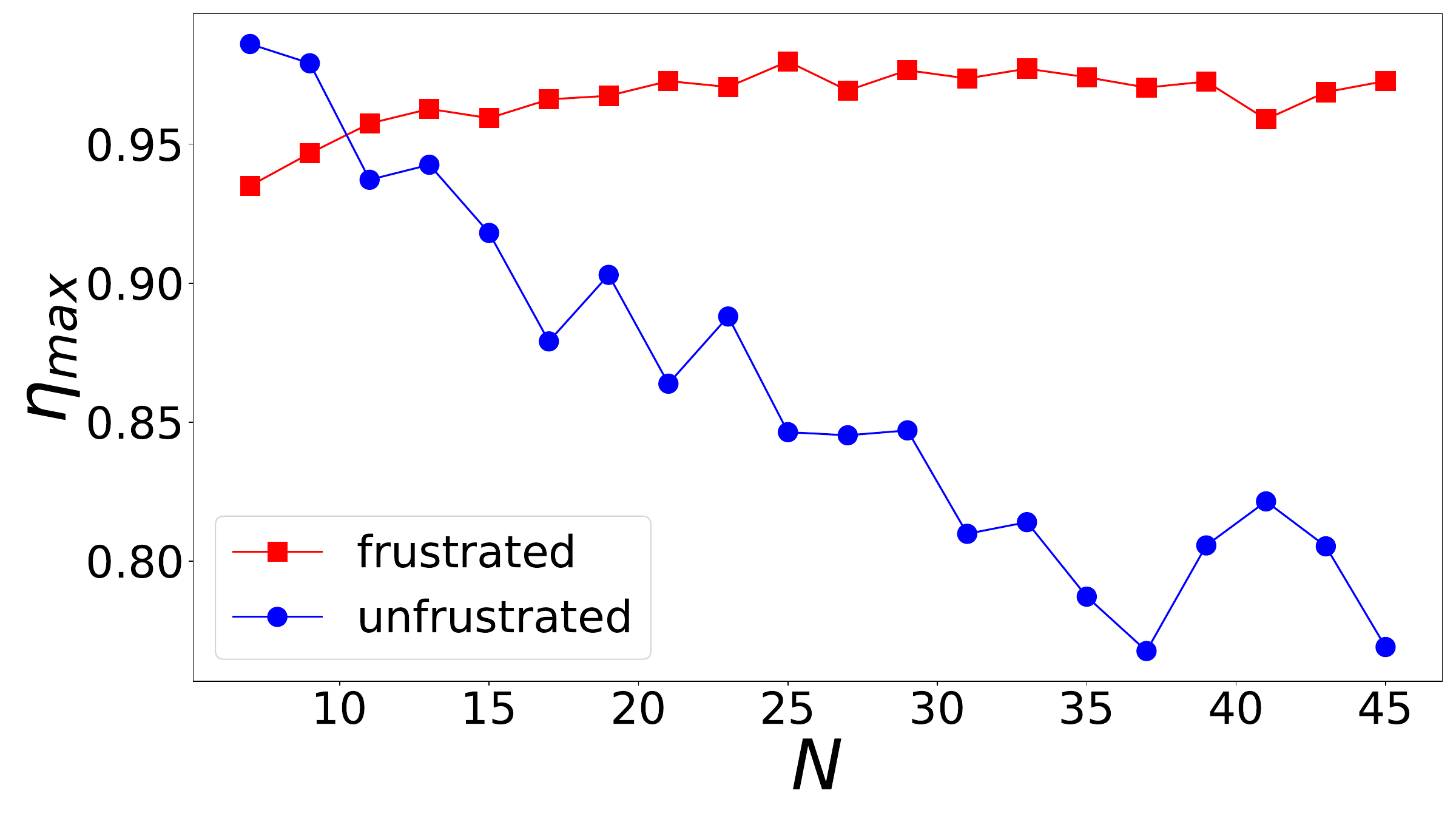}
\caption{\small Maximum value of $\eta$ {in Eq. \eqref{eq:eta}}, computed for $\tau \in (0,50)$, for the frustrated (red full dots) and frustrated (blue empty dots) Ising chain. Data are obtained for chains of size $N\in[7,45]$, for $h_0=0.751$ and fixed $h_1=0.7$.}
\label{fig:eta_N}
\end{figure}

Following the prescription of~\cite{Alla2004, Niedenzu2019} the lowest energy state that can be reached with (reversible) unitary process acting on the density matrix {$\tilde{\rho}(T)$ is its passive counterpart 
\begin{eqnarray}
\tilde{\rho}(T) =\sum_\ell \tilde{P}_{\ell}(\tau) |\epsilon_\ell\rangle \langle \epsilon_{\ell}|\;,
\end{eqnarray} 
where $\tilde{P}_{\ell}(\tau)$ are the eigenvalues of  $\rho_D$ rearranged in decreasing order ($\tilde{P}_{\ell}(\tau) \geq \tilde{P}_{\ell+1}(\tau)$).
The energy we can recover from the system via unitary operations can then  be computed in terms of the system ergotropy, i.e. the difference between
the input energy of $\rho_D$, and the mean energy of  $\tilde{\rho}_D$,
\begin{eqnarray}
W&=&\Tr(\rho(T) H_0)-\Tr(\tilde{\rho}_D H_0)\nonumber \\
&=& \sum_\ell (P_{\ell}(\tau) - \tilde{P}_{\ell}(\tau)) (\epsilon_\ell-\epsilon_0)\\
&=& E_{in}-\sum_\ell \tilde{P}_{\ell}(\tau) (\epsilon_\ell-\epsilon_0)=E_{in}-E_{loss}\;.\nonumber
\label{eq:wde}
\end{eqnarray}
The quantity $E_{loss}=\sum_\ell \tilde{P}_{\ell}(\tau) (\epsilon_\ell-\epsilon_0)$ represents the amount of energy that we cannot extract any more from the battery.
Since $E_{loss}$ is a positive quantity,
we have that, due to the non-unitary dynamics acting after the end of the charge phase, the work $W$ that we can extract from the battery is less than the energy $E_{in}$ stored in it.
To quantify how robust the QB is towards decoherence, we compute the ratio between the amount of work we can extract from it at time $T\gg\tau_2$ and the energy {initially} stored in the QB, i.e.
\begin{equation}
\label{eq:eta}
\eta=\frac{W}{E_{in}}. 
\end{equation}

The results obtained with a semi-analytical approach, see App.~\ref{ap:population}, are shown in Fig.~\ref{fig:eta_h}. 
The results are obtained by maximizing $\eta$ throughout the charging time $\tau$ in the interval $[0,50]$ in the unit of $1/\abs{J}$.
In the top panel, we depict the behavior of $\eta$ for a fixed value of $\Delta h=h_1-h_0$ as a function of $h_0$, while in the bottom one, we plot the result obtained keeping $h_0$ fixed and changing $\Delta h$.  

As we can see, in the top panel, well below the critical point $h_0=1$, the frustrated AFM battery is very resilient to the decoherence processes and, for a wide range of $h_0$,  $\eta$ is close to $1$ and well above $0.9$.
On the contrary, in the same range of parameters, the loss in the work extraction capability for an FM non-frustrated QB can go up to $80\%$. 
Moreover, in the bottom panel, the value of $\eta$ for the frustrated battery is strikingly close to $1$ in the whole range, while being considerably smaller for its non-frustrated counterpart.  

To understand the difference between the frustrated and non-frustrated systems, we have to consider the different characteristics of their energy spectrum. 
In the magnetically ordered phase of non-frustrated systems such as the one we are considering, the energy spectrum is characterized by two quasi-degenerate states separated from the first band of excited states by a finite energy gap. 
Conversely, in frustrated systems, the ground state belongs to a band made of $2N$ states whose width is related to the value of the external field. 
By comparing these behaviors, taking into account the definition of $E_{loss}$, it is easy to explain the different behavior. 
Indeed, in the case of non-topologically frustrated models, already the third term of the summation in the definition of $E_{loss}$ provides a non-negligible contribution to the loss of extractable energy and,
likewise, all the others that follow. 
Conversely, in frustrated systems, the contribution of the first $2N$ terms to $E_{loss}$, since all states belong to the same band, is small and decreases as $|h_0|$ decreases. 
This greatly reduces the loss of energy that can be extracted from the battery and, consequently, increases $\eta$.
However, when $|h_0|$ increases, the bandwidth of the frustrated model increases, reducing the value of $W$ and hence of $\eta$ while the gap of the ferromagnetic model narrows, resulting in an increment of $\eta$. These two different dependences on $|h_0|$ explain why, close to the quantum critical point, the performance of the two systems becomes comparable.
Moreover, since the number of states in the first band of the frustrated systems is proportional to the size of the system itself, it is natural to expect that the effect of reduction of the relative weight of $E_{loss}$ increases with $N$.
This expected behavior is confirmed by the results shown in Fig. \ref{fig:eta_N}. 
In varying the system size, the value of $\eta$ of the frustrated system remains approximately constant while it goes down with the system size for the non-frustrated model. }

\begin{figure}[t!]
\centering
\includegraphics[width=.9\columnwidth]{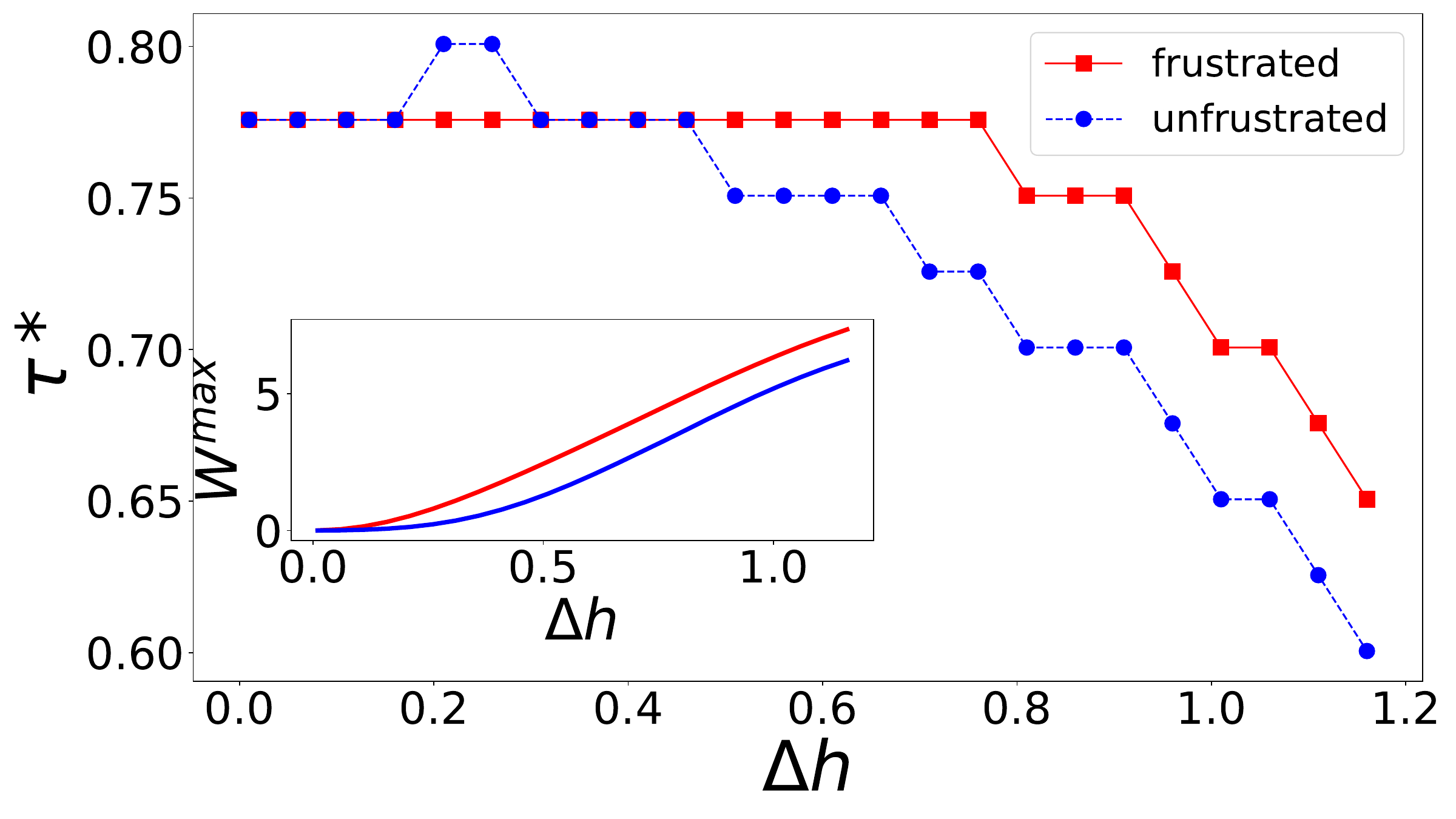}
\caption{\small Position of the first local time maximum $\tau^*$ and the corresponding value of $W$ (inset) for the frustrated (red) and unfrustarted (blue) Ising chain. Data are obtained for a chain of $N=25$ spins, for $h_0=0.001$ and $h_1-h_0 \in (0.01,1.2)$.}
\label{fig:peak_dh}
\end{figure}

{A further parameter that is useful in characterising the performances of the QB is the time needed to complete the charging process. Since for our model the amount of energy stored in the system is not a monotonic function of the duration of the charging process, we decided to define as charging time the one for which the first local maximum of the $E_{in}$ as a function of time is reached. 
This choice seems quite natural considering the necessity of having the charging time as short as possible. 
In Fig.~\ref{fig:peak_dh} we show the results of this analysis obtained varying $\Delta h_0$ for a fixed value of  $h_0$. 
From the figure, we observe that, regardless of the presence or the absence of topological frustration in the system, the charging time generally decreases with increasing $\Delta h$ while the energy stored in the system increases. 
This fact gives rise to a virtuous circle in which the time required for this storage decreases as the energy stored by the system increases. 
Moreover, for frustrated systems, the ratio $\eta$ always remains greater than $0.8$ and significantly higher than the one of the non-frustrated counterparts. 
This means that by increasing the jump in the external magnetic field it is possible to charge the battery more, faster, and with higher resistance to decoherence. 
While this behavior is valid for both frustrated and non-frustrated QBs, the data witness the fact that the performances of the first are always better than the ones of the second.}

\section{Batteries in the slow-charging regime}\label{sec:deco-charg} 
{The results {presented} so far were obtained under the hypothesis that the charging protocol was so fast that we can completely neglect any decoherence effects during it. 
However, such a hypothesis is quite strong and an analysis of what happens when the charging process is affected by decoherence is mandatory.}
Therefore, in this section, we study the performance of our QB model in the slow-charging regime where the fast decoherence time $\tau_1$ and the charging time $\tau$ are comparable. {To this end, we numerically integrate Eq. \eqref{eq:milburn} during the charging time up to $\tau$. } 
Even if, in this regime, the analysis is more complex, the basic concepts are the same as in the previous section, and we recover that, after the end of the charging process, waiting for a time $T\gg \tau_2$ the state of the QB reduces to a completely incoherent state of the form 
\begin{eqnarray}
\label{diagorhoT}
 \rho(T) \simeq \sum_{\ell} P'_{\ell}( \tau)  |\epsilon_\ell\rangle \langle \epsilon_{\ell}|\;,
 \end{eqnarray} 
where the populations $P'_{\ell}( \tau)$ are
\begin{eqnarray} \label{popuk} 
P'_{\ell} (\tau) &=&\sum_{k,k'}  \langle \epsilon_{\ell}| \mu_k\rangle \langle \mu_k|\epsilon_0\rangle
\langle \epsilon_0|  \mu_{k'}\rangle \langle \mu_{k'}|\epsilon_{\ell}\rangle \times \nonumber \\
&\times&  \exp[-\tfrac{(\mu_k-\mu_{k'})^2}{2\nu} \tau  -i (\mu_k-\mu_{k'})\tau ]\;, 
\end{eqnarray} 
(see Appendix~\ref{App:integral} for details) that correctly reduces to Eq.~(\ref{popuk}) when all the exponential decays can be neglected. 
{Note that, although the derivation of Eq. \eqref{eq:milburn} in \cite{Milburn1991} is not valid in the $\nu \to 0$ limit, we can take this limit of fast dephasing by instanteneously removing all off-diagonal contributions.}

\begin{figure}[t!]
\includegraphics[width=.9\columnwidth]{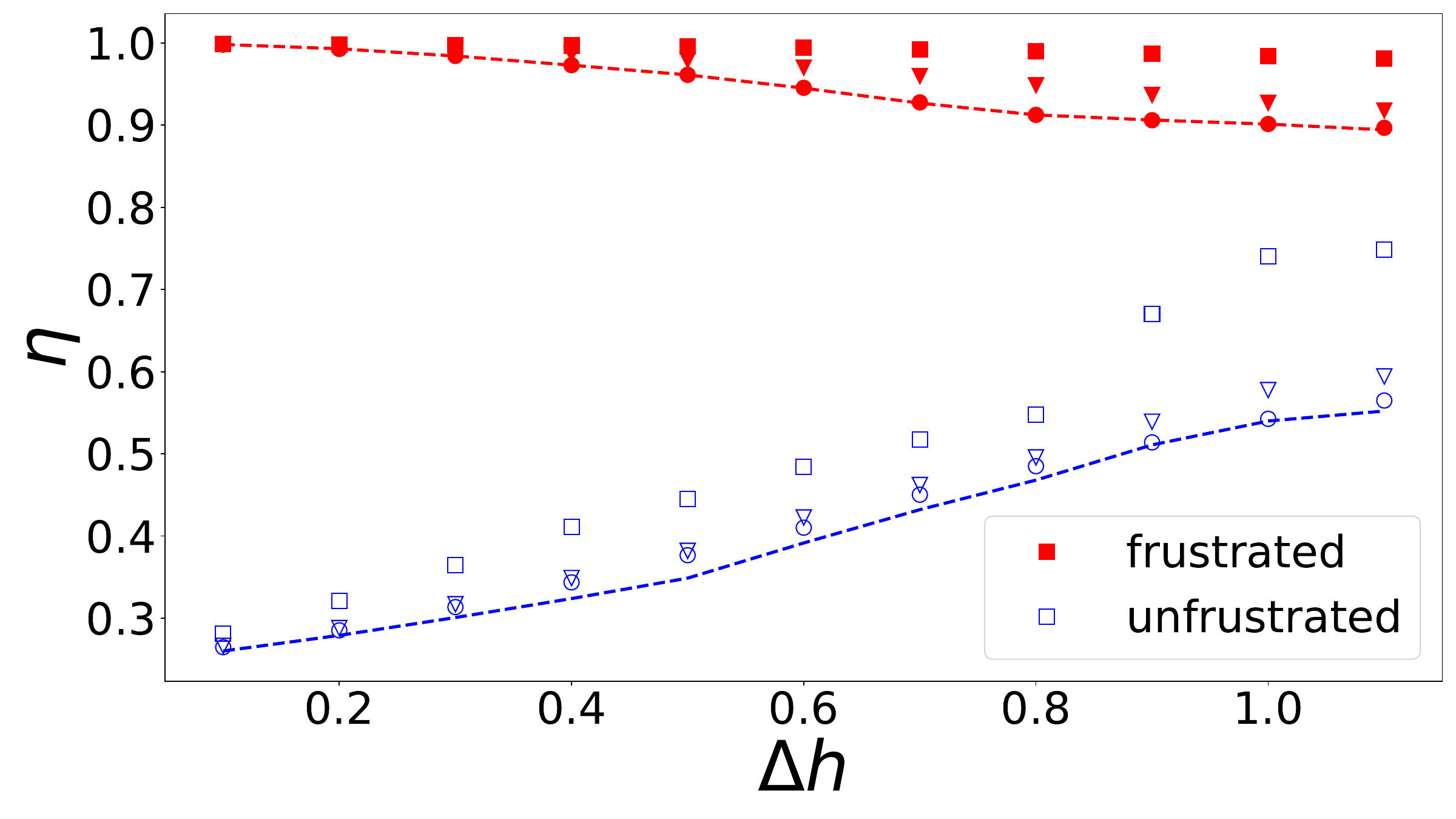}

\caption{\small Maximum in time of the robustness parameter $\eta$ for the frustrated (red) and non-frustrated (blue) for a decoherent charging protocol, for different values of the decoherence frequency: $\nu=10, 1, 0.1, 0$ (squares, triangles, circles, dashed line). $\nu=0$ corresponds to full decoherence, i.e. {instantaneous} convergence to the diagonal ensemble. Data are obtained for a chain of $N=15$ spins, for $h_0=0.001$ and $h_1-h_0 \in (0.02,1.1)$.}
\label{fig:deco_charge}
\end{figure}

Similarly to what was done in the previous section, we have compared the performances of the frustrated and non-frustrated QB models using the ratio $\eta$ and the velocity of charging.  
We show the outcomes of the analysis in Fig.~\ref{fig:deco_charge}. For several values of the decoherence frequency $\nu$, we charged the battery as a function of $\Delta h$, maximizing over time the robustness parameter $\eta=W/E_{in}$.
As expected, by decreasing the decoherence frequency $\nu$, hence making the decoherence stronger and faster in destroying the coherence of the QB state, $\eta$ decrease but does not disappear completely even in the limiting case  $\nu=0$ where all the off-diagonal elements of the density matrix are instantaneously suppressed soon after the quench of the external field from $h_0 \to h_1$. 
However, once again, the frustrated battery shows a higher robustness with respect to its non-frustrated counterpart. 
Even in the limiting case of $\nu=0$ (dashed line), the value of $\eta$ is above $0.9$ for a frustrated battery, while it drops below $0.5$ for the non-frustrated one. 
These drops in $\eta$ are more pronounced for larger values of the quench jump $\Delta h$. 
Indeed, for smaller quenches, the most populated state is still the ground state. Hence, at least in principle, one could still try to get as close as possible to the initial state when discharging the battery. 
However, this becomes more difficult when increasing $\Delta h$, as the number of excited states that are macroscopically populated increases, and therefore the loss of quantum coherence has a stronger impact on the value of the ergotropy and hence of $\eta$. 

\begin{figure}[t!]
\includegraphics[width=.9\columnwidth]{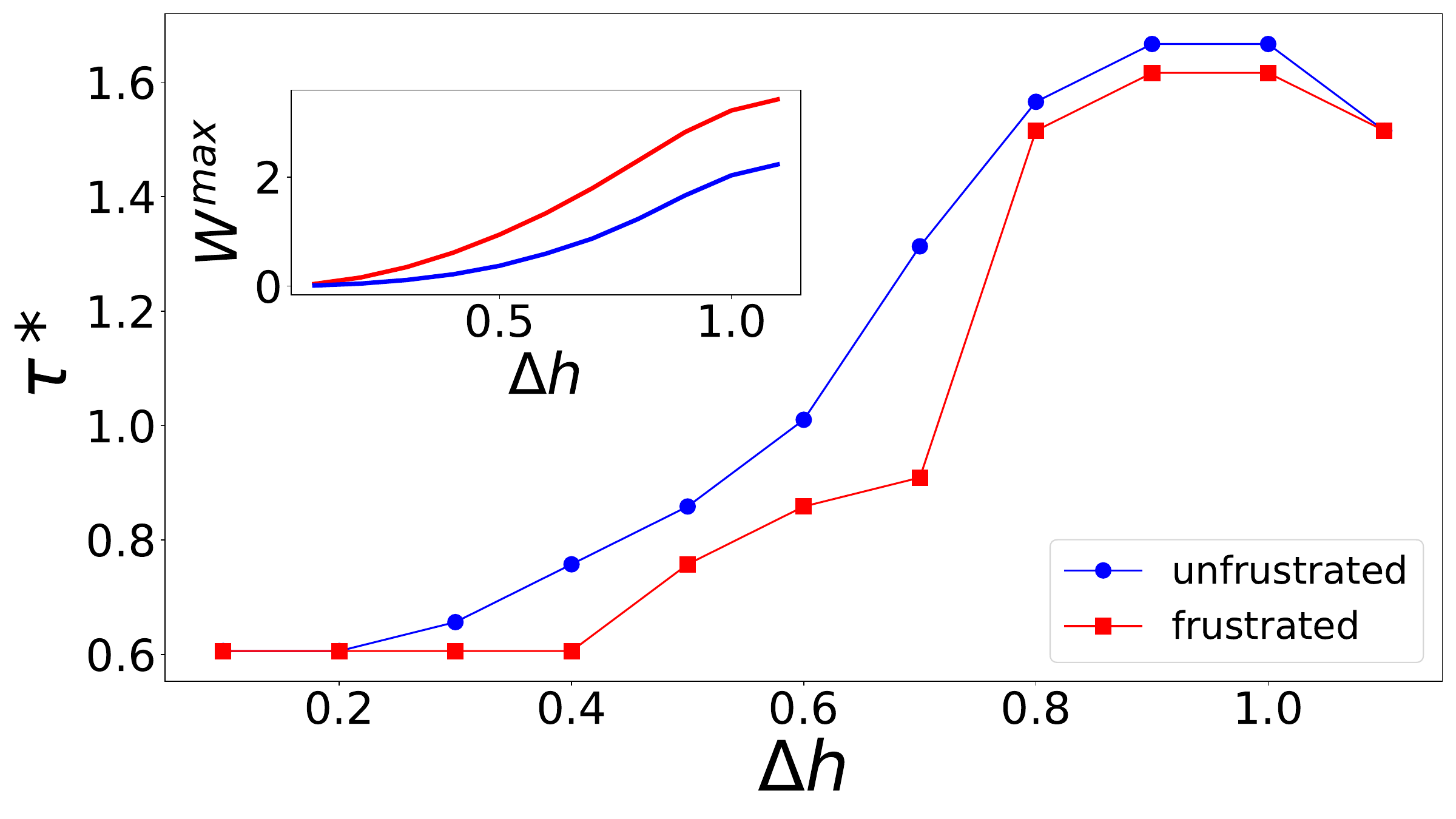}
\caption{\small Position of the first local time maximum $\tau^*$ and the corresponding value of $W$ for the frustrated (red) and unfrustarted (blue) Ising chain. Data are obtained for a chain of $N=15$ spins, for $h_0=0.001$ and $h_1-h_0 \in (0.01,1.2)$ and $\nu=1$.}
\label{fig:deco2}
\end{figure}

The decoherence also affects the charging time, making the charging slower for both the frustrated and the non-frustrated batteries. 
Therefore, we extend the analysis carried out in the previous section also to the case of the slow charging process. 
The results in Fig.~\ref{fig:deco2} confirm the fact that the charging processes for both frustrated and non-frustrated systems are comparable, but, the virtuous circle that we have seen in the fast charging regime has disappeared. {Indeed, while in the fast charging regime by increasing the quench amplitude $\Delta h$ we would increase the ergotropy of the battery and observe a reduction of the charging time $\tau$, in the slow charging regime it is still true that the ergotropy increases with the quench amplitude, but instead the charging time tends to increase, reducing the perfomances of the quantum battery.}  

\section{Discharging the battery}\label{sec:discharge}

{Up to now, we were mainly focused on the ergotropy of the system and how much it could be affected by the presence of an unavoidable non-unitary dynamic that continues to act even after the end of the charging process.  
However, such a quantity represents an upper bound of energy that can be extracted from a QB, which is hard to approach when this last is represented by a many-body system.
Therefore, in this section, we have decided to analyze a more realistic situation. 
We take into account a situation in which a QB, after ending the charging process and waiting a time $T\gg\tau_2$ such that its state can be considered completely incoherent, is made to interact with an ancillary two-level system. }
The Hamiltonian of the total system will therefore read as
\begin{equation}
\label{eq:HQBplusA}
H_W=J\sum_{k=1}^N \sigma^x_{k}\sigma^x_{k+1}-h\sum_{k=1}^N\sigma^z_{k} + \lambda H_{int} + \omega \sigma_S^z,
\end{equation}
where $\{\sigma_k^\alpha\}_{k=1}^N$ and $\{\sigma_S^\alpha\}$ ($\alpha=x,y,z$) are respectively the spin operators of the $k$-th site of the QB and the ancillary spin, while $\omega$ is the characteristic energy of the ancillary spin. 

{To simulate a realistic condition, we consider that only one of the spins of the battery directly interacts with the ancillary system. 
Moreover, for the same reason, we do not try to optimize the kind of interaction between the QB and the ancillary system, which can give rise to extremely hard-to-simulate interactions, but we directly take into account a realistic one as the hopping term.
Accordingly with these assumptions, $H_{int}$ reads}
\begin{equation}
\label{eq:Hint}
H_{int}=\sigma^+_1 \sigma^-_S+\sigma^-_1 \sigma^+_S,
\end{equation}
where $\sigma^\pm_1=\sigma^x_1\pm i \sigma^y_1$ and $\sigma^\pm_S=\sigma^x_S\pm i \sigma^y_S$ and the strength of the interaction is parametrized by $\lambda$.
Therefore, the ancillary spin is interacting with a single spin in the battery. 
The interaction that we have chosen breaks both the translational invariance and the parity symmetry of the battery, increasing the number of states accessible during the discharging process. 
Moreover, it can be experimentally realizable in Rydberg atom systems~\cite{Barredo2015}. 
Before going further, let us underline that, ideally, one would want an interaction term that commutes with the rest of the combined battery/system Hamiltonian. 
However, since our battery is a many-body system where the eigenstates are delocalized, this would require an interaction term that interacts with the battery as a whole. However, such interaction, even if theoretically achievable, would be unrealistic from an experimental point of view. 


In our simulation, we consider that, at time zero, the battery and the ancillary system are brought into contact by turning on the interaction in the equation. 
Before this, the two systems have been prepared.
As far as the QB is concerned, we have at first charged it with the unitary protocol, stopping at the time of the first peak in $\eta$, for $h_0=0.018$, $h_1=1.5$ and then let it relax to the diagonal ensemble.
On the contrary, the ancillary system is initialized in its lowest energy state $\rho_S$, i.e. spin down configuration, for $\omega=2|J|=2$.
This value of $\omega$ is chosen in such a way as to resonate with the band gaps of the battery spectrum, which for $h_0$ close to the classical point are proportional to $J$. 
The strength of the interaction between the spin and the battery was chosen small enough that the interaction does not carry too much energy into the system, but strong enough to allow for energy transfer. 
We established numerically that $\lambda=0.02$ is a good compromise that ensures that no appreciable energy is absorbed or released from the interaction term. 

Soon after $t=0$, the global system, initialized in the product state $\rho_B\otimes \rho_s$, is allowed to evolve under the action of global Hamiltonian $H_W$ and the energy starts to flow from the QB to the ancillary system.  
As for all systems, when some energy is provided to the ancillary spin, a part of it will be stored as work, while the rest will be dissipated as heat. 
Hence, a critical point is whether and how much of this energy can be seen as work performed by the QB on the ancillary spin. One way to reply to this question is to analyze the ratio $\kappa$ between the ergotropy $W_S$ acquired by the ancillary spin (that is initialized in a zero ergotropy state, i.e. its ground state) and its maximal ergotropy, i.e. 
\begin{equation}
\kappa=\frac{W_S}{2\omega}.
\end{equation}
In other words, $\kappa$ is the amount of energy that can be later used by the ancillary spin to perform some work.

\begin{figure}[t!]
\centering
\includegraphics[width=.99\columnwidth]{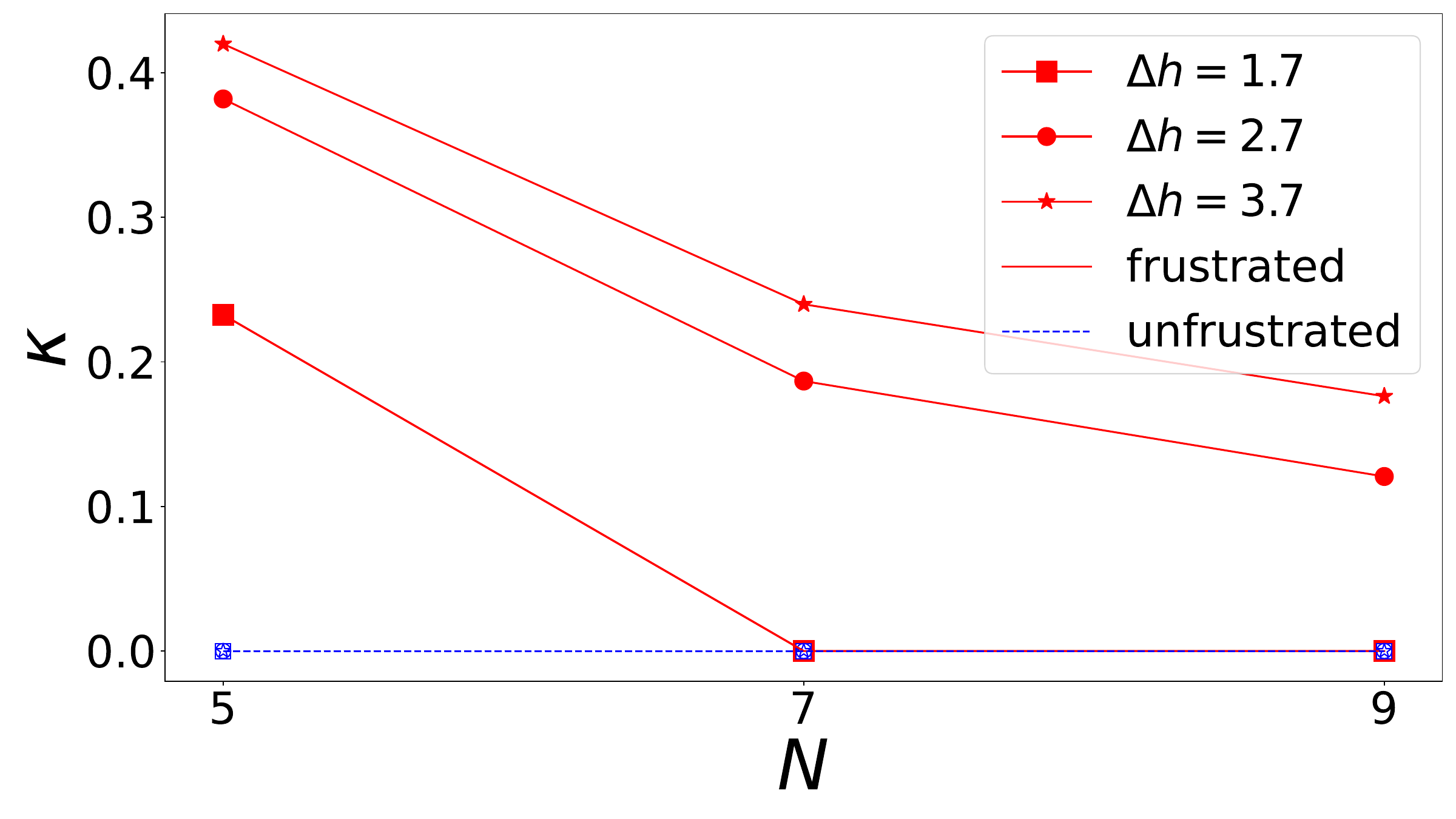}
\caption{\small Ergotropy charged in the ancillary spin from the frustrated battery (full red) and non-frustrated (empty blue) batteries of $N=5$, $7$, and $9$  spins. The results are obtained for $h_0=0.02$, $\omega=2$, $\lambda=0.02$ and $h_1=1.72$, $2.72$ and $3.72$.}
\label{fig:ex5_1}
\end{figure}

The results obtained through exact diagonalization for $\kappa$, for both the frustrated and non-frustrated battery, are shown in Fig. \ref{fig:ex5_1}, for different values of $\Delta h$ and of the size of the chain. 
The plot shows that none of the energy transferred from the non-frustrated battery is translated into ergotropy for the ancillary spin. 
On the contrary, the frustrated battery manages to charge the ancillary spin up to $42\%$ of its maximal ergotropy. 
This percentage decreases with increasing size of the battery, due to the very local nature of the interaction between the battery and the ancillary spin.

\section{Conclusion and discussion of results}\label{sec:conclusion}

{We propose a quantum battery based on a quantum many-body system, namely the quantum Ising chain, designing a cyclic charging protocol based on a global quench in the external magnetic field to store energy in the battery. We used different figures of merit to characterize the efficiency of such devices. In every case, we observed that the frustrated batteries present a very strong resistance to decoherence effects.} This remarkable result is related to the fact that the ground state of the frustrated system belongs to a gapless band, allowing for a more efficient energy extraction with respect to the non-frustrated models, where the presence of a finite gap between the ground state and the rest of the spectrum increases the energy of the final equilibrium state of the battery, therefore reducing the fraction of energy which is possible to extract. {Thus, our results show that topologically frustrated systems
can represent a much more efficient option for the realization
of a quantum battery with respect to their
non-frustrated counterparts.}

For the ergotropy, we tested the stability of these results by varying the parameters governing the model ($N$, $h_0$) and the charging protocol ($\Delta h$). In all the measured ranges we always observe a higher robustness to decoherence for the frustrated model. In the range of parameters that we considered, the charging time of the frustrated and non-frustrated batteries are comparable, even though for some values of the parameter we have observed shorter charging times for the non-frustrated one. However, even when the non-frustrated battery is charged a bit faster, the frustrated battery still possesses a higher ergotropy and a larger value of $\eta$. We expect these results to be valid even after increasing the system size. Moving towards the thermodynamic limit, one might expect that the value of $\eta$ might decrease since the system will start populating states in higher energy bands (this has to be better investigated). However, at the same time, the density of the states within a band will increase as the gaps tend to close as $N^{-2}$, and for the frustrated system, the degeneracy will always be larger than for the non-frustrated one. Therefore, also in the thermodynamic limit we expect the frustrated model to be more robust to decoherence than the non-frustrated one.

Moreover, we analyzed what happens in a discharging process in which we connect an ancillary spin to the battery. We defined a protocol that allows energy transfer from the battery to the spin and measured the level of charge acquired by the spin, measured as the fraction of its maximal ergotropy $\kappa$. The results show that the energy transferred from the non-frustrated battery is not translated into ergotropy for the ancillary spin, while, within the considered parameters, the frustrated battery charges the spin up to $42\%$ of the maximal ergotropy, which is $2\omega$. Therefore, once again we find that the performance of the frustrated battery, {due to internal correlations of the chain,} overcomes its non-frustrated counterpart. 

As a final remark, we would like to point out that spin models such as the 1D quantum Ising chain can be experimentally realized with Rydberg atoms. In the currently available experimental platforms, typical values of the couplings are $\Tilde{J}\approx \hbar \cdot 672$ MHz, $\Tilde{h}\approx \hbar \cdot 25$ MHz, and the system can be stabilized for times of the order of $20-70$ $\mu s$. The fastest decoherence time scale of the system can be estimated as the time it takes for the oscillating coherences to average out, i.e. $\tau_d\approx \hbar / (\Tilde{J}\Delta E)$, where $\Delta E$ is the largest dimensionless energy difference between the energy eigenstates populated after the quantum quench for the Ising chain described by the dimensionless Hamiltonian $H(1,\tilde{h}/\Tilde{J})$. Since $\Delta E$ is of order unity, the typical decoherence time will be of the order of a few tenths of nanoseconds. For the parameters mentioned above, we would have $\tau_d\approx 1.4$ ns. Since this time is considerably smaller than the time for which the system can be stabilized, decoherence effects might become relevant for a quantum battery realized on these platforms. Therefore, topologically frustrated quantum batteries, because of their high robustness to decoherence, might represent a valid alternative for the realization of these quantum devices.   

AGC and OM acknowledge support from the MOQS ITN programme, a European Union’s Horizon 2020 research and innovation program under the Marie Sk\l{}odowska-Curie grant agreement number 955479. 
SMG and FF, acknowledge support from the QuantiXLie Center of Excellence, a project co-financed by the Croatian Government and the European Union through the European Regional Development Fund – the Competitiveness and Cohesion (Grant KK.01.1.1.01.0004).
FF and SMG  also acknowledge support from the Croatian Science Foundation (HrZZ) Projects No. IP–2019–4–3321. OM acknowledges support from the Julian Schwinger Foundation.
VG and OM acknowledge financial support by MUR (Ministero dell’ Universit\`a e della Ricerca) through the PNRR MUR project PE0000023-NQSTI.

\appendix
\section{Solution of the Ising model: frustrated vs non-frustrated case}
\label{ap:Ising}

It is well known that the Ising model in equation~\eqref{eq:Ising} can be diagonalized exactly.
For the sake of simplicity, we limit our analysis to the case with $0\le h < 1$, but our results can be easily extended also to the other regions of parameters space. We will consider the case of an odd number of spins $N$ and periodic boundary conditions, i.e. we apply the frustrated boundary conditions FBC. In such a way we can easily switch from the frustrated to the non-frustrated model by setting $J=1$ or $J=-1$ respectively. In particular, in the first case, in presence of an antiferromagnetic interaction between nearest neighbors spins, frustration has been proven to produce non-trivial modifications to the ground-state properties of the model. 
The standard procedure prescribes a mapping of spin operators into fermionic ones \cite{Franchini17}, which are defined by the Jordan-Wigner transformation:
\begin{equation}
	\label{eq:JW-a}
	\sigma^-_j=\prod_{l<j} \sigma^z_l\psi_l^\dagger, \quad \sigma^+_j=\prod_{l<j} \sigma^z_l\psi_j, \quad \sigma^z_j=1-2\psi_j^\dagger\psi_j,
\end{equation}
where $\psi_l$ ($\psi^\dagger_l$) are fermionic annihilation (creation) operators.
In terms of such operators, taking into account the periodic boundary conditions and discarding constant terms, the Hamiltonian thus becomes
\begin{eqnarray}
	H\!&\!=\!&\!J\sum_{j=1}^{N-1}\!\bigg[\psi^\dagger_{j+1}\psi_j\!+\!\psi^\dagger_{j}\psi_{j+1}\!+\!\psi^\dagger_{j}\psi^\dagger_{j+1}\!+\!\psi_{j+1}\psi_j \bigg]  \\
	\! &\! +\! &\! 2h\sum_{j=1}^N\psi^\dagger_j\psi_j \!+\! -J\Pi^z \bigg[\psi^\dagger_{1}\psi_N\!+\!\psi^\dagger_{N}\psi_{1}\!+\!\psi^\dagger_{N}\psi^\dagger_1\!+\!\psi_{1}\psi_N\bigg].\nonumber
\end{eqnarray}
The latter expression is not quadratic itself, but reduces to a quadratic form in each of the parity sectors of $\Pi^z$. Therefore, it is convenient to write it in the form  
\begin{equation*}
	H=\frac{1+\Pi^z}{2}H^+\frac{1+\Pi^z}{2}+\frac{1-\Pi^z}{2}H^-\frac{1-\Pi^z}{2},
\end{equation*}
where both $H^\pm$ are quadratic. 
Hence we can bring the Hamiltonian into a free fermion form by means of two final steps. 
First, we perform a Fourier transform
\begin{equation}
	\psi_q=\frac{e^{-\imath \pi/4}}{\sqrt{N}}\sum_{j=1}^N e^{-\imath qj}\psi_j.
\end{equation}
It is worth noting that, due to the different quantization conditions, $H^\pm$ are defined on two different sets of fermionic modes, respectively $q\in\Gamma^-=\{\frac{2\pi n}{N}\}_{n=0}^{N-1}$ in the odd sector and $q\in\Gamma^+=\{\frac{2\pi}{N}(n+\frac{1}{2})\}_{n=0}^{N-1}$ in the even one.
Finally a Bogoliubov rotation in Fourier space 
\begin{equation}
	\label{eq:hi-a}
	b_q=\cos \theta_q \psi_q + \sin \theta_q \psi^\dagger_{-q}, 
\end{equation}
with momentum-dependent Bogoliubov angles 
\begin{equation}
	\label{eq:bogoliubov_angles-a}
	\theta_q=\frac{1}{2}\arctan\bigg(\frac{\sin q}{h+J\cos q}\bigg) \;\; q\neq0,\pi\;\;,\;\; \theta_{0,\pi}=0,
\end{equation}
leads to the Hamiltonians
\begin{subequations}
\begin{eqnarray}
\label{eq:ham-}
H^-\!&\!=\!&\!\!\!\!\!\!\!\sum_{q\in \Gamma^-/\{0\}}\!\!\!\!\!\!\Lambda(q)\bigg(b^\dagger_q b_q\!-\!\frac{1}{2}\bigg) +\epsilon(0)\bigg(b^\dagger_0 b_0\!-\!\frac{1}{2}\bigg)\\
\label{eq:ham+}
H^+\!&\!=\!&\!\!\!\!\!\!\!\sum_{q\in \Gamma^+/\{\pi\}}\!\!\!\!\!\!\Lambda(q)\bigg(b^\dagger_q b_q\!-\!\frac{1}{2}\bigg) +\epsilon(\pi)\bigg(b^\dagger_\pi b_\pi\!-\!\frac{1}{2}\bigg),
\end{eqnarray}
\end{subequations}
Here $b_q$ ($b^\dagger_q$) are the Bogoliubov annihilation (creation) fermionic operators.
The dispersion relation $\Lambda(q)$ for $q\neq0,\,\pi$ obeys
\begin{equation}
\label{eq:lambda}
\Lambda(q)=\sqrt{(h+J\cos q)^2+\sin^2q}, 
\end{equation}
while for the two specific modes $q=0\in\Gamma^-$ and  $q=\pi\in\Gamma^+$ we have 
\begin{equation}
\epsilon(0)=h+J, \quad \epsilon(\pi)=h-J.
\end{equation}
It is important to observe that, all fermionic modes are associated with a positive energy, but the $0$ and $\pi$ mode which carry negative energy respectively for $J=-1$ and $J=1$ since we have chosen $h<1$. 
Let us start by considering the FM case $J=-1$. In this case, for $0\le h < 1$ the only mode that carries negative energy is the $0$-mode in the odd sector, while all the modes in the even parity sector carry positive energy. {Therefore, in this case the ground state is given by a state with a $0$-mode populated $b^\dagger_0\ket{\emptyset^-}$ in the odd parity sector. One can observe that the vacuum states in each sector can be written in terms of the fermionic states $\ket{0,1}_q$ and of the Bogoliubov angles $\theta_q$, and read}
\begin{subequations}
	\label{eq:groundstate-a}
	\begin{eqnarray}
		\label{eq:groundstate-e-a}
		\ket{\emptyset^+}\!&\!=\!&\!\ket{0_\pi}\!\bigotimes_{q\in\Gamma_2^+}\!\left(\cos{\theta_q}\ket{0}_q\ket{0}_{-q}\!-\!\sin{\theta_q}\ket{1}_q\ket{1}_{-q} \right) \;\;\; \;\;\; \\
		\label{eq:groundstate-o-a}
		\ket{\emptyset^-}\!&\!=\!&\!\ket{0_0}\!\bigotimes_{q\in\Gamma_2^-}\!\left(\cos{\theta_q}\ket{0}_q\ket{0}_{-q}\!-\!\sin{\theta_q}\ket{1}_q\ket{1}_{-q} \right)\;\;\;\; \;\;
	\end{eqnarray}
\end{subequations}
where $\Gamma^+_2$ ($\Gamma^-_2$) is the subset of momenta $q \in \Gamma^+$ ( $q \in \Gamma^-$) that live in the interval $q\in(0,\pi)$. One can explicitely check that the ground-state energy is given by 
\begin{equation}
E_0^-=-\frac{1}{2}\sum_{q\in \Gamma^-}\Lambda(q).
\end{equation}

Let us now focus on the analysis of the AFM model obtained for $J=1$. {In this case we observe that the $\pi$-mode is the only one that can be characterized by a negative excitation energy.  
However, the $\pi$-mode exists only in the even parity sector where the addition of a single excitation is forbidden by the parity constraint.}

{For this reason, the lowest energy state of the even sector in this region is still its Bogoliubov vacuum $\ket{\emptyset^+}$.}
In the odd sector, states with a single excitation are allowed, but all fermionic modes hold positive energy and, it is easy to check that each state that can be defined in this sector has an energy greater than the one associated to $\ket{\emptyset^+}$. Despite this, the lowest admissible states in this sector, those with one occupied mode with momentum closest to $\pi$ (exactly $\pi$ is not possible because of the quantization rule of this sector) have an energy gap closing as $1/N^2$ compared to $\ket{\emptyset^+}$. The ground-state energy is given by
\begin{equation}
E_0^+=\Lambda(\pi)-\frac{1}{2}\sum_{q\in \Gamma^+}\Lambda(q).
\end{equation}
	

\section{Projection coefficients after a global quench}
\label{ap:population}
In this section we compute analytically the projection coefficients after a global quench from an Hamiltonian $H_0\equiv H(J,h_0)$ to $H_1\equiv H(J,h_1)$. Since the fermionic structure of the states is the same, the results hold both for the non-frustrated (FM) and frustrated (AFM) case. The intial state before the quench is considered to be the ground state of $H_0$. In the FM ($J=1$) case this can be given by
\begin{equation}
\ket{G_0^+}=\ket{\emptyset^+}_0,
\end{equation}
or
\begin{equation}
\ket{G_0^-}=b^\dagger_0\ket{\emptyset^-}_0=\ket{0}_0,
\end{equation}
depending on the parity sector, while in the AFM ($J=-1$) the ground state is always in the even sector and we have

\begin{equation}
\ket{G_0^+}=
\ket{\emptyset^+}_0.
\end{equation}

In both situations, since the global quench in the magnetic field preserves the translational invariance and parity of the model, after the quench the initial state will have non-vanishing projection only onto those eigenstates of $H_1$ with its same parity and momentum, i.e. states with zero momentum. {Moreover, since all of the eigenstates are constructed by addition of quasi-particles with a certain quasi-momentum $q$ to a fermionic vacuum, it turns out that the projections will be non-zero only onto those states where excitations are added in couples with opposite momentum, i.e. applying the operator $b^\dagger_q b^\dagger_{-q}$ to the ground-state.} Using simple combinatorics, one could hence easily understand that in a system with $N$ spins, starting from the intial states $\ket{\emptyset^+}$ or $\ket{0}$ the number of states with non-zero projections will be 
\begin{equation}
M=\sum_{l=0}^{\frac{N-1}{2}}\binom{\frac{N-1}{2}}{l}=2^{\frac{N-1}{2}}.    
\end{equation}

The projections can explicitly be computed by evaluating scalar products between different states. These are easily evaluated when the states are expressed in the fermion basis rather than in the Bogoliubov one as in \eqref{eq:groundstate-a}, since the fermionic operators are independent of the parameters of the hamiltonian, which will only enter the Bogoliubov angles. Using the notation 
\begin{equation}
\ket{\emptyset_k}=\cos \theta_k\ket{0}_k\ket{0}_{-k}-\sin \theta_k\ket{1}_k\ket{1}_{-k},
\end{equation}

we also have that
\begin{equation}
b^\dagger_k b^\dagger_{-k}\ket{\emptyset_k}=\sin \theta_k\ket{0}_k\ket{0}_{-k}+\cos \theta_k\ket{1}_k\ket{1}_{-k}.
\end{equation}
Therefore, because of the selection rules imposed by the global quench, we have only four possibilities for the scalar products after the quench:
\begin{equation}
\bra{\emptyset_k^{(1)}}\ket{\emptyset_k^{(0)}}=\cos\Delta_k,
\end{equation}
\begin{equation}
\bra{\emptyset_k^{(1)}}b^{\dagger(0)}_k b^{\dagger (0)}_{-k}\ket{\emptyset_k^{(0)}}=-\sin\Delta_k,
\end{equation}
\begin{equation}
\bra{\emptyset_k^{(1)}}b^{\dagger(1)}_{-k} b^{\dagger (1)}_{k}\ket{\emptyset_k^{(0)}}=\sin\Delta_k,
\end{equation}
\begin{equation}
\bra{\emptyset_k^{(1)}}b^{\dagger(1)}_{-k} b^{\dagger (1)}_{k}b^{\dagger(0)}_k b^{\dagger (0)}_{-k}\ket{\emptyset_k^{(0)}}=\cos\Delta_k.
\end{equation}
where $\Delta_k=\theta^{(1)}_k-\theta^{(0)}_k$. 

Finally, we can introduce the notation $\ket{P_0}=\prod_{p\in P_0} b^\dagger_p b^\dagger_{-p}\ket{G_0}$, to the describe a generic zero-momentum state, $P_0$ being a subset of $\Gamma^+\setminus\{\pi\}$ or $\Gamma^-\setminus\{0\}$ depending on the parity sector. With this in mind, the projection coefficient that we are looking for will take the form
\begin{equation}
\bra{Q_1}\ket{P_0}=\prod_{\substack{k_1\in \Gamma \setminus (Q_1\cup P_0 \cup \{0,\pi\}), \\ k_2  \in Q_1\cap P_0, \\ k_3 \in P_0\setminus Q_1,\\ k_4\in Q_1\setminus P_0} }\!\!\!\!\!\!\!\!\!\!\!\!\!\!\!\cos \Delta_{k_1} \cos \Delta_{k_2} (-\sin \Delta_{k_3}) \sin \Delta_{k_4}.
\end{equation}
These coefficients, for opportune choices of the quasi-momenta, will correspond to the $\bra{\epsilon_k}\ket{\mu_\ell}$ appearing in Eq. \eqref{popuk}. Therefore, knowing them allows us to compute the populations $P_\ell$.
\section{Formal integration of Eq.~(\ref{eq:milburn})} \label{App:integral} 
In case where the  system Hamilton $H$ is time independent, the ME~(\ref{eq:milburn})
admits analytical integration \cite{Milburn1991}.  To see this let us write $H$ as 
\begin{eqnarray}
H = \sum_{\epsilon} \Pi_{\epsilon} \epsilon \;,  
\end{eqnarray} 
where $\epsilon$ are the eigenvalues of such operator, and $\{ \Pi_{\epsilon}\}_{\epsilon}$ is  the set of orthogonal projectors which decompose the the Hilbert space of the system in the associated energy eigenspaces. Exploiting the fact that $\sum_{\epsilon} \Pi_{\epsilon} = \openone$, 
$\Pi_{\epsilon}\Pi_{\epsilon'}=\delta_{\epsilon,\epsilon'} \Pi_{\epsilon}$, one can then
verify that an explicit solution of~(\ref{eq:milburn}) is provided by 
\begin{equation} 
\rho(t) = 
\Phi^{(H)}_t [\rho(0)] = \sum_{\epsilon,\epsilon'} \Pi_\epsilon \rho(0) \Pi_{\epsilon'}
e^{ - \frac{(\epsilon -\epsilon')^2}{2\nu} t - i (\epsilon -\epsilon')t}\;,  
\end{equation} 
where $\Phi^{(H)}_t$ is the dynamical superoperator~\cite{Open} 
\begin{eqnarray}
\Phi^{(H)}_t [\cdots] = \sum_{\epsilon,\epsilon'} \Pi_\epsilon \cdots \Pi_{\epsilon'}
e^{ - \frac{(\epsilon -\epsilon')^2}{2\nu} t - i (\epsilon -\epsilon')t}\;. 
\end{eqnarray} 
Notice that for $t\gg \tau_2$, where $\tau_2$ is the long dephasing time identified in the main text, such evolution induce complete suppression of the off-diagonal terms that involves superpositions associated with energy eigenvectors of different eigenvalues, i.e. 
\begin{equation} \label{dephasing} 
\Phi^{(H)}_t [\cdots]\Big|_{\nu t\gg 1}  \quad \longrightarrow\quad  {\cal D}^{(H)} [\cdots] = \sum_{\epsilon} \Pi_\epsilon \cdots \Pi_{\epsilon}\;. 
\end{equation} 
For the model we are considering $H$ is equal to $H_1$ for $t \in ]0,\tau[$ and to $H_0$
for $t\geq \tau$. Accordingly we can write 
\begin{eqnarray}
\rho(t) = \left\{ \begin{array}{lr} 
\Phi_t^{(H_1)} [\rho(0)]\;, & \forall t \in ]0,\tau[\;, \\ \\ 
\Phi_{t-\tau}^{(H_0)}\left[ \Phi_{\tau}^{(H_1)}[\rho(0)]\right]\;, & \forall t\geq \tau\;,
\end{array} \right.
\end{eqnarray} 
which for  $t=T\geq \tau$ such that $T\gg \tau_2$, leads to 
\begin{eqnarray}
\rho(T) &\simeq &\label{dephasingH0} 
{\cal D}^{(H_0)}\left[ \Phi_{\tau}^{(H_1)}[\rho(0)]\right]\;,
\end{eqnarray} 
with ${\cal D}^{(H_0)}$ the dephasing map~(\ref{dephasing}) of $H_0$. 
Equation~(\ref{diagorhoT}) finally follows from~(\ref{dephasingH0}) observing that 
under the assumption the the initial state of the QB is the ground state of $H_0$, then 
all the eigenspaces involved in the writing of both of $\Phi_{\tau}^{(H_1)}$ and $\Phi_{t-\tau}^{(H_0)}$ only involves eigenspaces with zero momentum which turn out to be
non-degenerate (i.e. their associated projectors are all rank one). 
 

\begin{thebibliography}{99}
























\bibitem{Acin2018} A. Ac\'{\i}n, {\it et al.}, New Journal of Physics, {\bf 20}, 080201, (2018).

\bibitem{Riedel2017} M. F. Riedel, D. Binosi, R. Thew, and T. Calarco, Quantum Science and Technology, {\bf 2}, 030501, (2017).

\bibitem{Alicki2013} R. Alicki and M. Fannes,
Phys. Rev. E {\bf 87}, 042123 (2013).

\bibitem{Binder2015} F.C. Binder, S. Vinjanampathy, K. Modi, and J. Goold,
New J. Phys. {\bf 17}, 075015 (2015).


\bibitem{Campaioli2017} F. Campaioli, F.A. Pollock, F.C. Binder, L. C\'eleri, J.
Goold, S. Vinjanampathy, and K. Modi, Phys. Rev.
Lett. {\bf 118}, 150601 (2017).

\bibitem{Ferraro2018} 
D. Ferraro, M. Campisi, G.M. Andolina, V. Pellegrini,
and M. Polini, Phys. Rev. Lett. {\bf 120}, 117702 (2018).


\bibitem{Andolina2018}  G. M. Andolina, D. Farina, A. Mari, V. Pellegrini, V. Giovannetti, and M. Polini, 
Phys. Rev. B, vol. {\bf 98}, 205423 (2018).

\bibitem{Farina2019} D. Farina, G. M. Andolina, A. Mari, M. Polini, and V. Giovannetti, 
Phys. Rev. B,  {\bf 99}, 035421 (2019).

\bibitem{Andolina2019b} 
G.M. Andolina, M. Keck, A. Mari, M. Campisi, V. Giovannetti, and M. Polini, Phys. Rev. Lett. {\bf 122}, 047702
(2019).



\bibitem{Hovh2013} 
K. V. Hovhannisyan, M. Perarnau-Llobet, M. Huber, and A.Ac\'{\i}n,
Phys. Rev. Lett., {\bf 111}, 240401  (2013).


\bibitem{Ghera2020} 
S. Gherardini, F. Campaioli, F. Caruso, and F. C. Binder, 
Physical Review Research, {\bf 2}, 013095 (2020).

\bibitem{Rosa2020} D. Rosa, D. Rossini, G. M. Andolina, M. Polini, and M. Carrega, 
J. High E. Physics {\bf 2020}, 67 (2020).






\bibitem{Tirone2021} S. Tirone, R. Salvia, and V. Giovannetti, Phys. Rev.
Lett. {\bf 127}, 210601 (2021).
\bibitem{Tirone2022} S. Tirone, R. Salvia, S. Chessa, and V. Giovannetti,
	arXiv:2211.02685  [quant-ph].
\bibitem{Tirone2023a} S. Tirone, R. Salvia, S. Chessa, and V. Giovannetti,
	arXiv:2304.01270 [quant-ph].
	\bibitem{Tirone2023b} S. Tirone, R. Salvia, S. Chessa, and V. Giovannetti,
	arXiv:2305.16803 [quant-ph].


\bibitem{Rodri2022} 
R. R. Rodriguez, B. Ahmadi, G. Suarez, P. Mazurek, S. Barzanjeh, and P. Horodecki, 
Eprint Arxive: quant-ph2207.00094, (2022).


\bibitem{Pirmo2019} 
F. Pirmoradian and K. M{\o}lmer, 
Physical Review A {\bf 100}, 43833 (2019).

\bibitem{Mazzoncini2022} F. Mazzoncini, V. Cavina, G. M. Andolina, P. A. Erdmann,
and V. Giovannetti, Phys. Rev. A {\bf 107}, 032218 (2022). 

\bibitem{Erdman2022}  P. A. Erdmann, G. M. Andolina, V. Giovannetti, and 
F. No\'e, arXiv:2212.12397 [quant-ph].







\bibitem{Andolina2019a} 
G.M. Andolina, M. Keck, A. Mari, V. Giovannetti, and
M. Polini, Phys. Rev. B {\bf 99}, 205437 (2019).

\bibitem{Wang2020} 
Z. Wang, {\it et al.}, Phys. Rev. Lett. {\bf 124}, 013601 (2020).

\bibitem{Stock2017} 
A. Stockklauser, P. Scarlino, J.V. Koski, S. Gasparinetti,
C.K. Andersen, C. Reichl, W. Wegscheider, T. Ihn, K.
Ensslin, and A. Wallraff, Phys. Rev. X {\bf 7}, 011030 (2017). 
\bibitem{Samk2018} 
N. Samkharadze, G. Zheng, N. Kalhor, D. Brousse, A. Sammak, U.C. Mendes, A. Blais, G. Scappucci, and
L.M.K. Vandersypen, Science {\bf 359}, 1123-1127 (2018).

\bibitem{Haroche2012} S. Haroche, Rev. Mod. Phys. {\bf 85}, 1083 (2013).

\bibitem{Zhang2019} 
Y.-Y. Zhang, T.-R. Yang, L. Fu, and X. Wang, Phys. Rev. E {\bf 99}, 052106 (2019).
\bibitem{Crescente2020a} A. Crescente, M. Carrega, M. Sassetti, and D. Ferraro,
New J. Phys. {\bf 22}, 063057 (2020).
\bibitem{Crescente2020b}  A. Crescente, M. Carrega, M. Sassetti, and D. Ferraro,
Phys. Rev. B {\bf 102}, 245407 (2020).
\bibitem{Crescente2020c} A. Crescente, D. Ferraro, M. Carrega, and M. Sassetti,
Phys. Rev. Research {\bf 4}, 033216 (2022).
\bibitem{Dou2022a}  F.-Q. Dou, Y.-Q. Lu, Y.-J. Wang, and J.-A. Sun, Phys.
Rev. B {\bf 105}, 115405 (2022).
\bibitem{Dou2022b}  F.-Q. Dou, H. Zhou, and J.-A. Sun, Phys. Rev. A {\bf 106},
032212 (2022).
\bibitem{Zhao2022}
F. Zhao, F.-Q. Dou, and Q. Zhao, Phys. Rev. Research {\bf 4},
013172 (2022).

\bibitem{Rossini2019} D. Rossini, G. M. Andolina, and M. Polini, 
Phys. Rev. B, {\bf 100}, 115142
(2019).
\bibitem{Rossini2020} D. Rossini, G. M. Andolina, D. Rosa, M. Carrega, and
M. Polini, 
Phys. Rev. Lett.  {\bf 125}, 236402 (2020).

\bibitem{Quach2022} 
J.Q. Quach, K.E. McGhee, L. Ganzer, D.M. Rouse, B.W.
Lovett, E.M. Gauger, J. Keeling, G. Cerullo, D.G. Lidzey,
and T. Virgili, Science Advances {\bf 8}, eabk3160 (2022).

\bibitem{Monsel2020}
J. Monsel, M. Fellous-Asiani, B. Huard, and A. Auff`eves,
Phys. Rev. Lett. {\bf 124}, 130601 (2020).
\bibitem{Maffei2021}  M. Maffei, P.A. Camati, and A. Auff`eves, Phys. Rev.
Research 3, {\bf L032073} (2021).

\bibitem{Oppe2002} J. Oppenheim, M. Horodecki, P, Horodecki, and R.
Horodecki, Phys. Rev. Lett. {\bf 89}, 180402 (2002).



\bibitem{Carrega2020} 
M. Carrega, A. Crescente, D. Ferraro, and M. Sassetti, 
New J. Phys. {\bf 22}, 083085 (2020).
\bibitem{Bai2020}
 S.-Y. Bai and J.-H. An,
 Phys. Rev. A  {\bf 102}, 060201 (2020).
 \bibitem{Tabesh2020}  F. T. Tabesh, F. H. Kamin, and S. Salimi, 
 Phys. Rev. A  {\bf 102}, 052223 (2020).
\bibitem{Ghosh2021}
 S. Ghosh, T. Chanda, S. Mal, and A. Sen(De), 
 Phys. Rev. A {\bf 104}, 032207 (2021).
 \bibitem{Santos2021} 
A. C. Santos, 
Phys. Rev. E  {\bf 103}, 042118 (2021).
\bibitem{Zaka2021}
S. Zakavati, F. T. Tabesh, and S. Salimi, 
Phys. Rev. E  {\bf 104}, 054117 (2021).
\bibitem{Landi2021}
G. T. Landi, 
Entropy {\bf 23}, 10.3390/e23121627 (2021).
\bibitem{Morrone2023}
D. Morrone, M. A. C. Rossi, A. Smirne, and M. G. Genoni, 
Quantum Sci. Technol. {\bf 8}, 035007 (2023).
\bibitem{Sen2023}
K. Sen and U. Sen, 
arXiv:2302.07166 [quant-ph]

\bibitem{Liu2019} 
J. Liu, D. Segal, and G. Hanna, 
The Journal of Physical Chemistry C {\bf 123}, 18303 (2019).
\bibitem{Santos2019} 
A. C. Santos, B. Cakmak, S. Campbell, and N. T. Zinner, 
Phys. Rev. E {\bf 100}, 032107 (2019).
\bibitem{Quach2020} 
J. Q. Quach and W. J. Munro, 
Phys. Rev. App.  {\bf 14}, 024092 (2020).
\bibitem{Santos2020} 
A. C. Santos, A. Saguia, and M. S. Sarandy, 
 Phys. Rev. E {\bf 101}, 062114 (2020).
 
\bibitem{Liu2021}
J. Liu and D. Segal, 
arXiv:2104.06522 [quant-ph].
\bibitem{Arjmandi2022}
M. B. Arjmandi., H. Mohammadi, and A. C. Santos, 
Phys. Rev. E {\bf 105}, 054115 (2022).


\bibitem{Alla2004} 
A.E. Allahverdyan, R. Balian, and T.M. Nieuwenhuize,
Europhys. Lett. {\bf 67}, 565 (2004).

\bibitem{Niedenzu2019} 
W. Niedenzu, M. Huber, and E. Boukobza, 
Quantum {\bf 3}, 195 (2019).

\bibitem{Maric2020_induced}
V. Mari{\'c}, S. M. Giampaolo and F. Franchini, 
"Quantum phase transition induced by topological frustration", Communications Physics \textbf{3}, 220 (2020).

\bibitem{Maric2020_destroy}
V. Mari{\'c}, S. M. Giampaolo and F. Franchini, 
"The frustration of being odd: how boundary conditions can destroy local order",
New Journal of Physics, \textbf{22}, 083024 (2020).

\bibitem{Catalano2022}
A. G. Catalano, D. Brtan, F. Franchini and S. M. Giampaolo, "Simulating continuous symmetry models with discrete ones", Physical Review B, \textbf{106}, 125145 (2022).
{
\bibitem{Maric2022_fate} V. Mari{\'c}, S. M. Giampaolo, and Fabio Franchini, "Fate of local order in topologically frustrated spin chains", Physical Review B {\bf 105}, 064408 (2022).

\bibitem{Giampaolo2019}
S. M. Giampaolo, F. B. Ramos and F. Franchini,
"The Frustration of being Odd: Universal area law violation in local systems", Journal of Physics Communication {\bf 3}, 081001 (2019).

\bibitem{Maric2022_nature} 
V. Mari\'c, G. Torre, F. Franchini, and S. M. Giampaolo, "Topological Frustration can modify the nature of a Quantum Phase Transition", SciPost Physics \textbf{12}, 075 (2022). 

\bibitem{Torre2022}
G. Torre, V. Marić, D. Kuić, F. Franchini, and S. M. Giampaolo, "Odd thermodynamic limit for the Loschmidt echo",
Physical Review B, \textbf{105}, 184424 (2022)

\bibitem{Odavic2022}
J. Odavi\'c, T. Haug, G. Torre, A. Hamma, F. Franchini, and S. M. Giampaolo, "Complexity of frustration: a new source of non-local non-stabilizerness", arXiv:2209.10541 (2022).
} 


\bibitem{ZENO} 
B. Misra, and E.C.G. Sudarshan, J. Math. Phys. {\bf 18}, 756 (1977). 

\bibitem{Open} F. Petruccione and H.-P. Breuer "The Theory of Open Quantum Systems",
(Oxford University Press, USA, 2007). 

\bibitem{Milburn1991}
G. J. Milburn, "Intrinsic decoherence in quantum mechanics", Physical Review A, \textbf{44}, 5401 (1991)

{
\bibitem{Baumgratz2014}
T. Baumgratz, M. Cramer, and M., B. Plenio, "Quantifying coherence", Physics Review Letters \textbf{113}, 140401 (2014).
}

\bibitem{Barredo2015}
D. Barredo, H. Labuhn, S. Ravets, T. Lahaye, A. Browaeys, and C. S. Adams, "Coherent Excitation Transfer in a Spin Chain of Three Rydberg Atoms",
Physiscal Review Letters, \textbf{114}, 113002 (2015)

\bibitem{Breuer}
H. P. Breuer, F. Petruccione, {\it The theory of open quantum systems}, Oxford University Press (2007).


\end{thebibliography}
\end{document}